\documentclass[prx,nofootinbib,citeautoscript,eqsecnum,10pt,longbibliography,notitlepage]{revtex4-2}
\pdfoutput=1

\usepackage{graphicx}
\usepackage{dcolumn}
\usepackage{color}
\usepackage{amsmath}
\usepackage{tabularx,graphicx}
\usepackage{epstopdf}
\usepackage{latexsym}
\usepackage{amssymb}
\usepackage{amsmath}
\usepackage{colortbl}
\usepackage{psfrag}
\usepackage{bbm,bm,array,physics}
\usepackage{titlesec}
\usepackage{dsfont}
\usepackage{xcolor}
\usepackage[tight]{subfigure}

\usepackage[papersize={8.5in,11in}]{geometry}

\definecolor{darkblue}{rgb}{0.,0.,0.4}
\definecolor{darkred}{rgb}{0.5,0.,0.}
\definecolor{BlueViolet}{RGB}{138,43,226}
\definecolor{SkyBlue}{RGB}{30,144,255}
\definecolor{DarkGreen}{RGB}{0,100,0}
\usepackage[pdftex,colorlinks=true,linkcolor=darkblue,citecolor=blue,urlcolor=darkred]{hyperref}
\def\*#1{\mathbf{#1}} 
\def \nn{\nonumber \\}
\def\ntl{\tilde{n}} 
 
\def\t{\mathrm} 
\def\lsb{\left[} 
\def\rsb{\right]} 
\def\lrb{\left(} 
\def\rrb{\right)} 
\def\nn{\nonumber \\}
\def\td#1{\tilde{#1}} 
\def\mcn{\mathcal{N}} 
\def\efano{E_{\text{Fano}}}

\def\gtin{\tilde{g}^{\text{i}}}
\def\gtout{\tilde{g}^{\text{o}}}

\def\jbessel{J_{n-m} \Bigg( \frac{V_1}{\hbar\omega} \Bigg)}
\def\rynp{r_{yn}^{+}}
\def\rynm{r_{yn}^{-}}
\def\rymp{r_{ym}^{+}}
\def\rymm{r_{ym}^{-}}
\newcommand\pmtx[1]{\begin{pmatrix}#1\end{pmatrix}} 

\begin{document}

\title{Floquet transmission in Weyl/multi-Weyl and nodal-line semimetals through a time-periodic potential well}

\author{Sandip Bera$^1$}
\author{Sajid Sekh$^2$}
\author{Ipsita Mandal$^{2,3}$}

\affiliation{$^1$Department of Physics, University of Toronto, 60 St. George Street, Toronto, Ontario,
	Canada M5S 1A7\\
	$^2$Institute  of  Nuclear  Physics,  Polish  Academy  of  Sciences,  31-342  Krak\'{o}w,  Poland\\
	$^3$Max-Planck Institute for the Physics of Complex Systems,
	N\"{o}thnitzer Str.~38, 01187, Dresden, Germany
}

\begin{abstract}
In mesoscopic physics, the application of a time-periodic drive leads to novel transport behaviour, which is absent in the static regimes. Here we consider a quantum pumping protocol, such that the quasiparticles of Weyl/multi-Weyl and nodal-line semimetals are subjected to a time-periodic rectangular potential well. The presence of an oscillating potential of frequency $\omega$ creates equispaced Floquet side-bands with spacing $\hbar \omega$. As a result, a Fano resonance is observed when the difference in the Fermi energy (i.e., the energy of the incident quasiparticle), and the energy of one of the (quasi)bound state levels of the well, coincides with the energy of an integer number of photons (each carrying energy quantum $\hbar \omega$, equal to the side-band spacing). Using the Floquet theory and the scattering matrix approach, in the zero-temperature non-adiabatic pumping limit, we find characteristic Fano resonance patterns in the transmission coefficients, which depend on the nature of the dispersion. The inflection points in the pumped shot noise spectra also serve as a proxy for the corresponding Fano resonances. Therefore, we also numerically evaluate the pumped shot noise. Finally, we correlate the existence of the Fano resonance points to the (quasi)bound states of the well, by explicitly calculating the bound states of the static well (which are a subset of the bound states of the driven system). Since we consider semimetals with anisotropic dispersions, all the features observed depend on the orientation of the potential well. We believe that our results will serve as a guide for future experiments investigating quantum transmission in nonequilibrium settings.
\end{abstract}

\maketitle

\tableofcontents

\section{Introduction} 
\label{sec:intro}

Quantum pumping is a viable candidate for generating directed bias-less
charge/spin currents through nanoscale devices, and therefore, has been a subject of extensive research in the field of mesoscopics. The first proposal of quantum pumping was put forward in the seminal work of Thouless in 1983 \cite{thouless83}, where the author showed that quantized transport of electrons takes place under a slowly-varying adiabatic potential. A few years later, quantum pumping was first realized in quantum dot systems by Gossard \textit{et al} \cite{switkes99}. At present, there exist several platforms to realize quantum pumping protocol in mesoscopic systems, such as quantum dot driven by microwave excitations \cite{oosterkamp98}, Gigahertz pumping \cite{blumenthal07}, and time-dependent gate voltages \cite{dicarlo03}. Only in the low-frequency limit (i.e., scattering by a slowly-modulated potential), we can apply the adiabatic approximation \cite{brouwer98,Moskalets04}, beyond which the pumping becomes non-adiabatic \cite{kaestner08,kim2004}. It is not possible to have single-parameter pumps in the adiabatic regime, where current is generated by driving just a single parameter, as the pump current vanishes in those situations \cite{jose2011,brouwer98}. Therefore, it becomes necessary to consider the non-adiabatic regime \cite{PhysRevLett.90.210602,PhysRevB.70.155326} to sustain a single-parameter current generation, and implement arbitrary driving frequencies that extend beyond the adiabatic regime.

Since pumping is associated with an ac gate voltage, it generates nonequilibrium shot noise features that carry various signatures of physical processes occurring inside the well in the mesoscopic regime. Of course, such information is washed away in the dc current obtained by time-averaging. Shot noise stems from current fluctuations due to the discreteness of the electrical charge \cite{beenakker03}. While a random and independent emission of electrons follows Poisson distribution in the shot noise spectrum, correlations among electrons reduce the noise below the Poissonian value. Therefore, the shot noise spectrum can be used to detect the nature of scattering, the amount of transferred charge, or the extent of entanglement. It can also serve as a proxy for resonance patterns in the transmission spectra, as we will explain below.

In the transmission spectrum of a driven potential well, the Fano resonance arises due to the interaction of the spatially localized states (i.e., the bound states) and the propagating modes. It is often interpreted as an interference effect of the electron wavefunctions along different quantum paths \cite{Zhu15}, leading to a resonance peak/dip in the transmission spectrum. Previous studies \cite{reichl199,Zhu15,Dai2014} on two-dimensional (2D) electron gas and graphene systems show the existence of Fano resonances in the meV ranges. Recently, such studies have been extended to include 2D pseudospin-1 Dirac-Weyl quasiparticles \cite{Zhu17} and quadratic band-touching semimetals \cite{Bera2021} (which encompass both the 2D and 3D versions). 
These examples include systems which are characterized by isotropic dispersions around the band-crossing (nodal) point. However, in addition to the most familiar examples of Dirac and Weyl nodes, semi-Dirac~\cite{pickett09} and multi-Weyl semimetals \cite{liu2017predicted,lundgren2014thermo,fang2012multi,pickett09} have been identified, which have a mix of linear and higher-order dispersions depending on the directions.
Due to their anisotropic dispersions, they exhibit contrasting electromagnetic properties~\cite{cho16,dietl08}, in comparison to the conventional Dirac and Weyl materials with purely linear dispersions. The anisotropy is also expected to play a significant role in several transport characteristics, such as quantum tunnelling effects \cite{PhysRevB.101.085410,ips-aritra,Khokhlov2018}, thermopower \cite{ips-kush}, chiral photocurrent \cite{ips-photocurrent}, circular dichroism \cite{sajid-cd}, Magnus Hall effect \cite{sajid_magnus22}, and magneto-transport \cite{ips-serena,ipsita-shivam}.

The physics of Floquet scattering is based on photon-assisted tunnelling, where the electrons tunnel through a potential well driven with frequency $\omega$, and exchange energy quanta (in units of $\hbar \omega$) in the process. In the Floquet scattering model, the oscillating modulation introduces Floquet ``side-bands'' in the resulting dispersion $E$, with ``quasienergies'' $ E + n \hbar\omega$, where $n \in \mathbb{Z}$ represents the order of the Floquet side-band. For an incoming electron with energy $E_F$, a Fano resonance shows up in the transmission spectrum when the difference in $E_F$ and one of the (quasi)bound state energy levels of the well (denoted by $E_b$) equals the energy of an integer number of photons (each carrying energy $\hbar \omega$ equal to the side-band spacing). The incident electron then emits photon(s) and drops to the bound state energy level. Alternatively, an electron already occupying a (quasi)bound state can absorb photon(s), and jump to the incident energy level or one of the Floquet channels $E_F + n \hbar\omega$. The resonance can be identified by the presence of a sharp peak or dip in the transmission spectrum, or alternatively, by an inflection point in the pumped shot noise. We use the Floquet scattering matrix \cite{reichl199,Zhu15,Dai2014,Zhu17,Bera2021,Araujo2021} (i.e., the S-matrix) formalism in the non-adiabatic limit to calculate both the transmission and the shot noise spectra.

In earlier studies \cite{reichl199,Zhu15,Dai2014,Zhu17,Bera2021}, Floquet scattering has been investigated for a variety of isotropic nodal-point semimetals. But such studies for anisotropic systems like nodal-line and nodal-point semimetals have been missing, which motivates this work. For systems with band-crossing points, we consider multi-Weyl semimetals featuring a linear dispersion along one direction (let us label this as the $z$-direction, without any loss of generality), and a quadratic/cubic dispersion in the plane perpendicular to it (i.e., the $xy$-plane). It can be shown that a multi-Weyl semimetal harbours a topological charge $J$, whose magnitude is higher than that of a Weyl semimetal with unit topological charge (i.e., $J=1$) --- $J=2$ at a double-Weyl node (e.g., $\mathrm{HgCr_2Se_4}$~\cite{Gang2011} and $\mathrm{SrSi_2}$~\cite{hasan_mweyl16}), and $J=3$ at a triple-Weyl node (e.g., transition-metal monochalcogenides~\cite{liu2017predicted}). In addition, we consider the nodal-line semimetals, where the band-touching occurs along a nodal ring \cite{cheng-nodal,Khokhlov2018}. Such dispersions appear in materials like $\mathrm{Cu_3PdN}$ \cite{xiao_nodal15}, $\mathrm{ZrSiS}$ \cite{fu_nodal19}, and $\mathrm{Mg_3Bi_2}$~\cite{chang_nodal19}.

The paper is organized as follows: We first review the Floquet formalism, scattering matrix theory, and the definition of shot noise in Sec.~\ref{sec:formalism}. We introduce the model Hamiltonians for Weyl/multi-Weyl and nodal-line semimetals in Sec.~\ref{sec:mweyl} and \ref{sec:nodalline}, respectively. These subsections also contain the computational details and the final plots, and additionally, we compare our numerical results with the previous studies for other semimetals. In Sec.~\ref{sec:boundstates}, we correlate the existence of the Fano resonances for each case to the energies of the bound states. Finally, we end with a summary and outlook in Sec.~\ref{sec:summary}. 

\begin{figure*}[]
	\centering
	\includegraphics[width = 0.75 \textwidth]{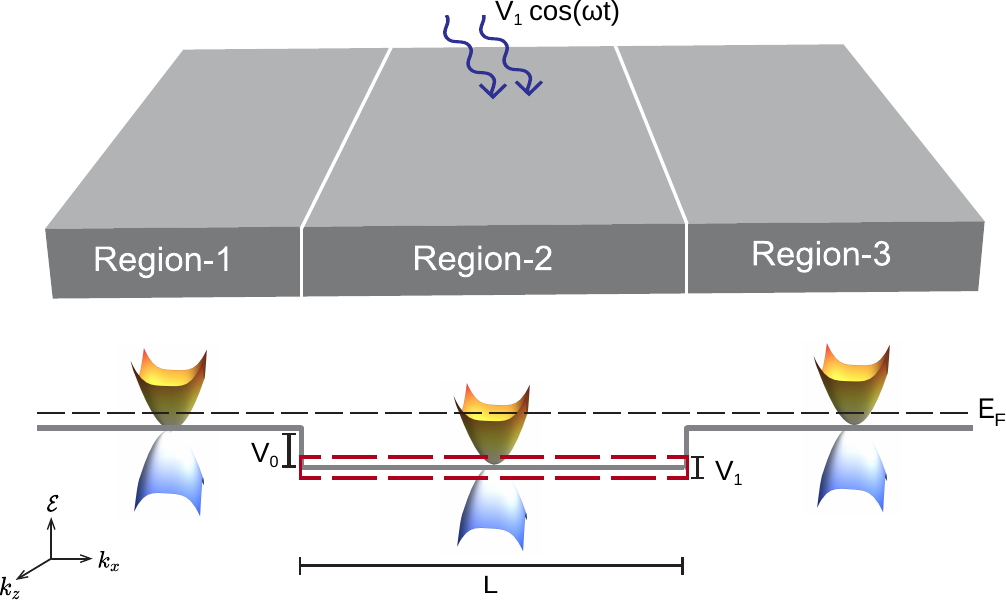}
	\caption{\label{fig:schematic}
		The schematic diagram for the tunnelling of quasiparticles in a double-Weyl ($J=2$) semimetal, subjected to a time-periodic potential well along the $x$-axis. The propagation of the wavefunction along the $x$-axis can be divided into three regions: region-1 is the region of incidence and reflection,  
		region-2 represents the potential well of length $L$ and magnitude $V_0$, and region-3 is the region of transmission. A harmonic drive of amplitude $V_1$ and frequency $\omega$ modifies the magnitude of the well periodically, which is shown by the red dashed lines. Outside the potential well, the multi-Weyl cone is filled up to the Fermi level $E_F$, as indicated by the black dashed line.}
\end{figure*}

\section{Floquet scattering and pumped shot noise} 
\label{sec:formalism}

The rectangular time-periodic potential well extending along the $a$-axis [where $a$ is one of the three mutually perpendicular axes $(x,y,z)$ in the Cartesian Coordinate system] is given by the function
\begin{align}
	\label{eqpotential}
	V(a,t)= \begin{cases} 
		-V_0+V_1\cos(\omega t)  &\t{ for } -L/2\le a \le L/2 \\
		0 &\text{ otherwise}
	\end{cases} \, ,
\end{align} 
where $L$ and $V_0$ are the length and magnitude of the depth of the well, respectively. The well is assumed to be infinite and homogeneous along the directions transverse to the $a$-axis, which means that for all practical purposes, the well has a sufficiently large cross-sectional width $W$, resulting in the conservation of the momenta in those directions. In our set-up, $V_1$ is the amplitude of the time-dependent drive with periodicity $\tau $, and $\omega = 2\pi/\tau$ is the frequency (cf. Fig.~\ref{fig:schematic}). All of these parameters can be controlled by ac fields in semiconducting devices~\cite{platero2004}. 

For a quasiparticle with wavefunction $\Psi(\*r,t)$, propagating in the $a$-direction, the time-dependent Schr\"{o}dinger equation implies
\begin{align} 
	\label{eq:schro_eq}
	i\, \hbar \, \partial_t \Psi(\*r,t) 
	= \left[ \mathcal{H}(-i\,\boldsymbol{\nabla})
	+ V(a,t) \right ] \Psi(\*r,t) \, ,
\end{align}
where the Hamiltonian $\mathcal{H}(-i\,\boldsymbol{\nabla})$ represents the bandstructure of a generic semimetal in the position space. Using the Floquet formalism, we can write the solution of Eq.~(\ref{eq:schro_eq}) as
\begin{align}
	\Psi(\*r,t) = \sum_{n=-\infty}^{\infty} \, e^{-i\,E_n\,t/\hbar}  \, e^{i\,k_b \,b}
	\, e^{i \,k_c \,c} \, \psi(a,t) \, ,
\end{align}
where $E_n = E_F + n\hbar\omega$ is the $n^{th}$ channel Floquet quasienergy (with $n\in \mathbb{Z}$), $E_F$ is the Fermi level, $\psi(a,t)$ obeys the periodicity $\psi(a,\, t+\tau) = \psi(a,\, t)$, and $b$ and $c$ refer to the Cartesian axes perpendicular to $a$. Since we have assumed conservation of momenta along the transverse directions (i.e., along the $b$- and $c$-axes), we have used a plane wave ansatz $e^{i\,k_b \,b}\, e^{  i\,k_c \,c}$ (with $k_b$ and $k_c$ real) along these directions. To find the wavefunction, we need to consider the three regions (as shown in Fig.~\ref{fig:schematic}): $a<-L/2$ (region-1), $-L/2 \leq a \leq L/2$ (region-2), $a > L/2$ (region-3). We parametrize the wavefunction piecewise in each of these regions with the appropriate $k_a$-momentum, and then impose the condition that the wavefunction must be continuous at the boundaries. Since we are dealing with two-band semimetals in this paper, we write the wavefunction as a two-component spinor $\psi(a,t) = \Big( \psi_{n,1}(a,t) 
\, \, \psi_{n,2}(a,t) \Big)^T$. The components can be written as follows \cite{reichl199,jose2011,Bera2021}:
\begin{align} 
\label{eq:psi}
\pmtx{\psi_{n,1}(a,t) \\ \psi_{n,2}(a,t)} =
\begin{cases} 
		A_n^{\t{i}}(t) \,e^{i \, k_n a} \pmtx{ {f^{\text{i}}} _{n1} \\  {f^{\text{i}}} _{n2}} + 
		A_n^{\t{o}}(t) \,e^{-i \, k_n a} \pmtx{ {f^{\text{o}}} _{n1} \\  {f^{\text{o}}} _{n2}}
		&\text{ for } a< -L/2 \\  & \\ 
		\sum \limits _{m=-\infty}^\infty
		\lsb 
		\alpha_m (t)\,e^{i \,  q_m a} \pmtx{ {\tilde{f}^{\text{i}}} _{m1} \\  {\tilde{f}^{\text{i}}} _{m2}}
		+  \beta_m (t)\,e^{-i \,  q_m a} \pmtx{ {\tilde{f}^{\text{o}}} _{m1} \\  {\tilde{f}^{\text{o}}} _{m2}} 
		\rsb
		J_{n-m}\left(\frac{V_1}{\hbar \omega}\right) \, \Theta(E_m+V_0) & \\ 
		+
		\sum \limits _{m=-\infty}^\infty
		\lsb
		\alpha_m (t)\,e^{i \,  q_ma} \pmtx{\gtin_{m1} \\ \gtin_{m2}}
		+  \beta_m (t)\,e^{-i \,  q_m a} \pmtx{\gtout_{m1}\\ \gtout_{m2}} 
		\rsb 
		J_{n-m}\left( \frac{V_{1}}{\hbar \omega}\right) \, \Theta(-E_m-V_0)
		&\text{ for }  -L/2 \le a \le L/2   \\ & \\
		B_n^{\t{i}}(t) \,e^{-i \, k_n a} \pmtx{ {f^{\text{o}}} _{n1} \\  {f^{\text{o}}} _{n2}} + 
		B_n^{\t{o}}(t) \,e^{i \, k_n a} \pmtx{ {f^{\text{i}}} _{n1} \\  {f^{\text{i}}} _{n2}}
		&\text{ for } a> L/2 \\  
	\end{cases},
\end{align}
where $J_n(x)$ is the $n^{th}$ order Bessel function of the first kind.
The coefficients $A_{n}^{\t{i}}(t)$ and $A_{n}^{\t{o}}(t)$ represent the amplitudes of the incoming and outgoing waves in region-1, and $B_{n}^{\t{i}}(t)$ and $B_{n}^{\t{o}}(t)$ represent the same for region-3. Similarly, $\alpha_m$ and $\beta_m$ denote the wavefunction amplitudes inside the potential well (i.e., region-2). The wavevectors $k_n$ and $q_m$ denote the components of the momentum along the $a$-axis, in the regions outside and inside the potential well, respectively. Finally, the symbols $\left \lbrace f, \tilde{f}, \tilde{g} \right \rbrace$ indicate the components of the spinors, and their explicit forms depend on the bandstructure and the direction of propagation under consideration.

Imposing the continuity conditions, we derive a set of equations linear in the amplitudes, which we express in a matrix form as
\begin{align}
	\label{eqsmatrix}
	s(E_n, E_{\ntl})= \sqrt{\frac{\Re[k_n]}{\Re[k_{\ntl}]}} \, \mathcal{S}_{n\ntl} \,,\quad
	\pmtx{ A_n^{\t{o}} \\  B_n^{\t{o}}} =
	\sum \limits_{\ntl=-\infty}^{\infty} \, \mathcal{S}_{n\ntl} \,. 
	\pmtx{ A_{\ntl}^{\t{i}} \\ B_{\ntl}^{\t{i}}}.
\end{align}
The matrix $\mathcal{S}_{n\ntl}$ encodes the probability amplitude that an electron in the $ {\tilde n}^{\rm{th}}$ channel is scattered to the $ n^{\rm{th}}$ channel. For calculating the unitary scattering matrix (or S-matrix), we should only consider the propagating modes, and discard the non-propagating (i.e., evanescent or decaying) modes with $E_n<0$, since the latter do not contribute to the probability current density. In a generic situation, the matrix $s$ takes the form:
\begin{align}
	s(E_n, E_{\ntl})= 
	\pmtx{s_{LL}(E_n, E_{\ntl})& s_{LR}(E_n, E_{\ntl}) \\ 
		s_{RL}(E_n, E_{\ntl})& s_{RR}(E_n, E_{\ntl})} =
	\pmtx{r_{n {\ntl}} & {\tilde t}_{n {\ntl}}  \\
		t_{n {\ntl}} & {\tilde r}_{n {\ntl}} } ,
\end{align}
where $s_{\alpha\beta}(E_n, E_{\ntl}) = \sqrt{\Re[k_n]/\Re[k_{\ntl}]} \, [\mathcal{S}_{n\ntl}]_{\alpha\beta}$, with $\alpha,\beta \in \lbrace L,R \rbrace $ denoting the ``left'' (abbreviated by ``$L$'') or the ``right'' (abbreviated by ``$R$'') lead. Additionally, for a pair of quasienergies $ E_n$ and $ E_{\ntl} $, $s_{\alpha\beta}(E_n, E_{\ntl})$ is an $n\times \ntl$ matrix, and composed of $\left \lbrace r_{n\ntl},  \,t_{n\ntl},
\,\td r_{n\ntl}, \,\td t_{n\ntl} \right \rbrace $. Here, $ r_{n\ntl}$ and $t_{n\ntl}  $ are the reflection and transmission amplitudes for an electron incident from the left, and propagating between the $\ntl^{\rm{th}}$ and $n^{\rm{th}}$ Floquet channels. On the other hand, $ \tilde r_{n\ntl}$ and $  \tilde t_{n\ntl} $ indicate the same, but for an electron incident from the right. To obtain the S-matrix from $s_{\alpha\beta}$, the indices $(n, \ntl) \in [0, \infty)$ (i.e., restricted to non-negative values), since it should include only the propagating modes. This matrix then represents the quantum mechanical amplitude for an electron with energy $E_{\tilde{n}}$ to enter the potential well region through lead $\beta$, absorb (for $n-\tilde{n}>0$) or emit (for $n-\tilde{n}<0$) $ |n-\tilde{n}| \hbar \omega $ photon quanta, and finally leave through lead $\alpha$ with energy $E_n$.
For the case of an electron that is incident from the left with Fermi energy $E_F$, we have $\ntl=0$, and $s_{\alpha\beta}$ reduces to a column matrix. Then the total transmission and reflection coefficients are given by
\begin{align}
	T= \sum \limits_{n=0}^{\infty} \, |t_{n0}|^2= |s_{RL}(E_{n},E_{F})|^2 \text{ and }
	R= \sum \limits_{n=0}^{\infty} \, |r_{n0}|^2
	=|s_{LL}(E_{n},E_{F})|^2 \,,
\end{align} 
respectively.

In mesoscopic systems, the application of a time-dependent drive produces a phase-coherent ac current. The noise properties of this current are of great interest, mainly because of two reasons: (i) it can help in our understanding of quantized charge transport, which may lead to quantized pumping \cite{avron01,kamenev00}; and (ii) the signature of noise contains nonequilibrium features that are not present in the time-averaged current. The noise has two components: thermal noise and shot noise. Here, we consider the zero temperature limit, and hence the thermal noise vanishes. The zero-temperature zero-frequency pumped shot noise is given by the expression \cite{Dai2014,Zhu15,Zhu_2011,Zhu10,Moskalets04}
\begin{align}
	\label{eqshotnoise}
	& \mcn_{\alpha\beta}(E_F)
	= \frac{e^2}{h} \, \int_0^\infty \, dE \, \sum_{\gamma,\delta=L,R} \; 
	\sum_{m,n,p=-\infty}^\infty 
	{\mathcal M}_{\alpha\beta\gamma\delta}(E,E_m,E_n,E_p) \, \frac{\lsb f_0(E_n) - f_0(E_m) \rsb^2}{2}, \nn
	& \text{where } {\mathcal M}_{\alpha\beta\gamma\delta}(E,E_m,E_n,E_p)  = 
	s^*_{\alpha\gamma}(E,E_n) \, s_{\alpha\delta}(E,E_m) \,
	s^*_{\beta\delta}(E_p,E_m) \, s_{\beta\gamma}(E_p,E_n) \, ,
\end{align}
which thus has the dimensions of $e^2/h$ times energy.
The zero-frequency shot noise corresponds to the noise measured in a time long enough compared to all intrinsic inverse-frequency scales as well as the pump time period $\tau$ \cite{Moskalets04}. It contains information about the energy exchanged during the scattering processes. The number of absorbed or emitted photon quanta is counted by the integers $(m,n,p)$, and $f_0$ is the Fermi-Dirac distribution. Hence, if we consider two quasiparticles propagating in the channels $\gamma$ and $\delta$, with energies $E_n$ and $E_m$, respectively, scattering takes place if $(E_n-E_m) $ is zero or an integer multiple of $ \hbar\omega$. After scattering, the two scattered quasiparticles transition to channel $\alpha$ with energy $E$, and channel $\beta$ with energy $E_p = E+p\hbar\omega$. Note that the components of $\mcn_{\alpha\beta}$ are related among themselves by the conditions $\mcn_{LL}=\mcn_{RR}=-\mcn_{RL}=-\mcn_{LR}$, which arise due to the particle flux conservation \cite{Zhu15}. Therefore, we will consider only $\mcn_{LL}$ in the rest of the paper, without any loss of generality. Since the inflection points in the pumped shot noise (which represent the resonance points of the transmission spectrum) may not very prominent, we also consider the differential shot noise $\partial \mcn_{LL} / \partial E_F$, where these features are magnified offering better visibility.

\section{Transport characteristics} 
\label{sec:results}

In this section, we numerically investigate the role of a time-dependent potential well in the transmission characteristics of the quasiparticles of Weyl/multi-Weyl and nodal-line semimetals. For simplicity, we use the natural units by setting $\hbar=e=c=1$. In our numerics, the Floquet side-band cutoff is set to $N=2$, so that $T = \sum^N_{n=0} |t_{n0}|^2$, and we have ensured that $N>V_1/(\hbar\omega)$. In the shot noise equation, each of $(m,n,p)$ runs from $0$ to $N$, while the upper limit of the integration in Eq.~\eqref{eqshotnoise} is truncated at $E_F+N\hbar\omega$.

\begin{figure}[] 
	\centering
	\subfigure[]{\includegraphics[width= 0.305 \textwidth]{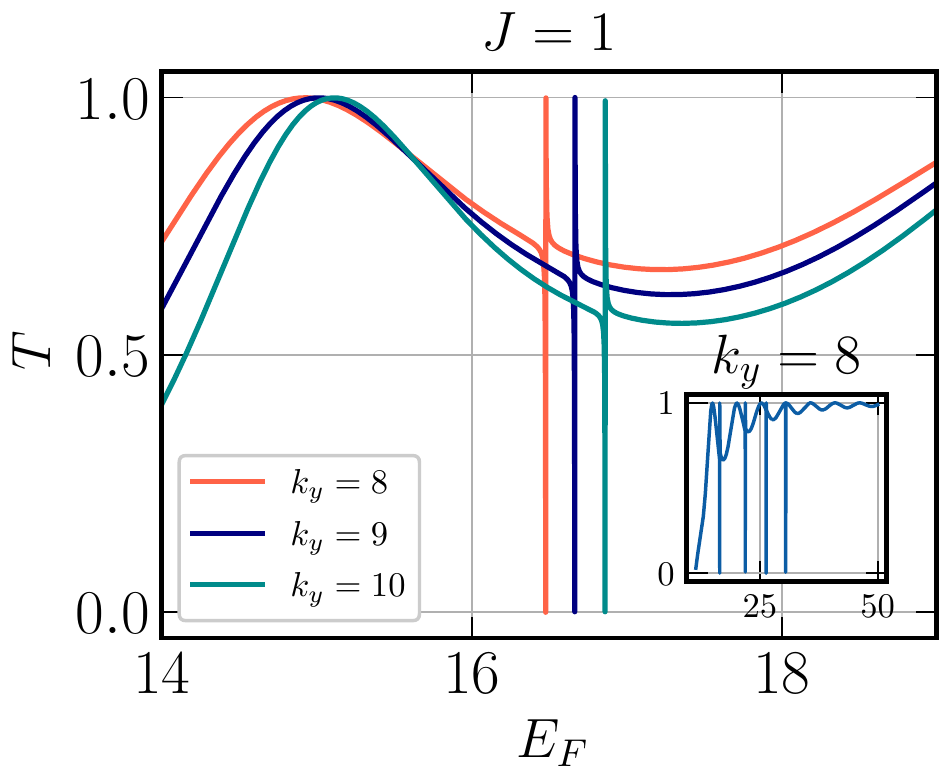}}\quad
	\subfigure[]{\includegraphics[width= 0.31\textwidth]{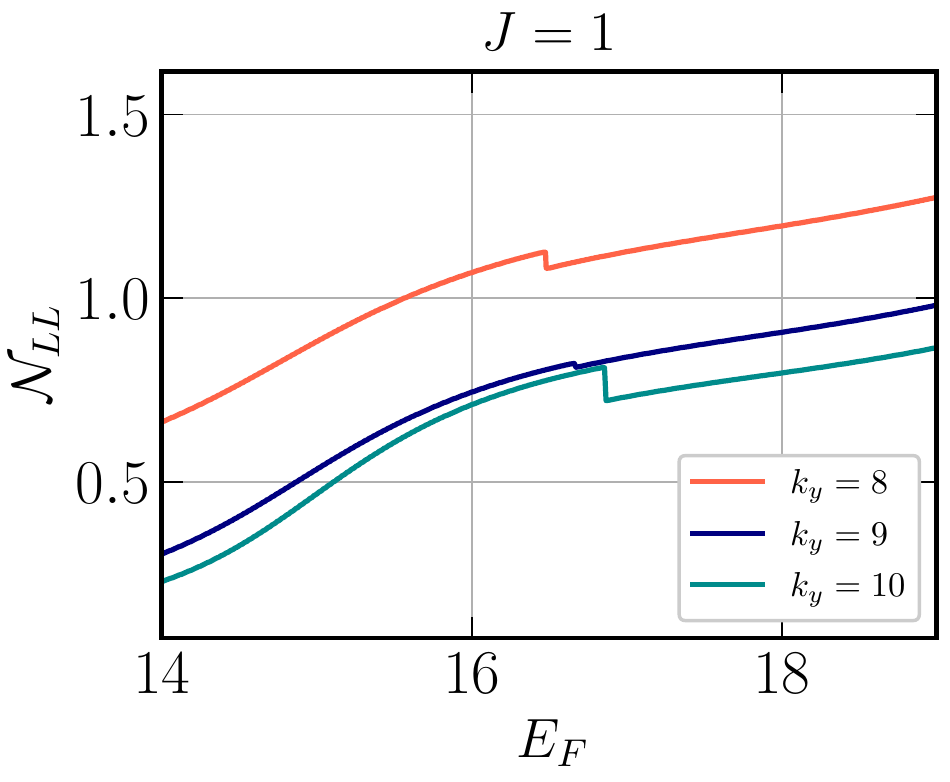}}\quad
	\subfigure[]{\includegraphics[width= 0.31\textwidth]{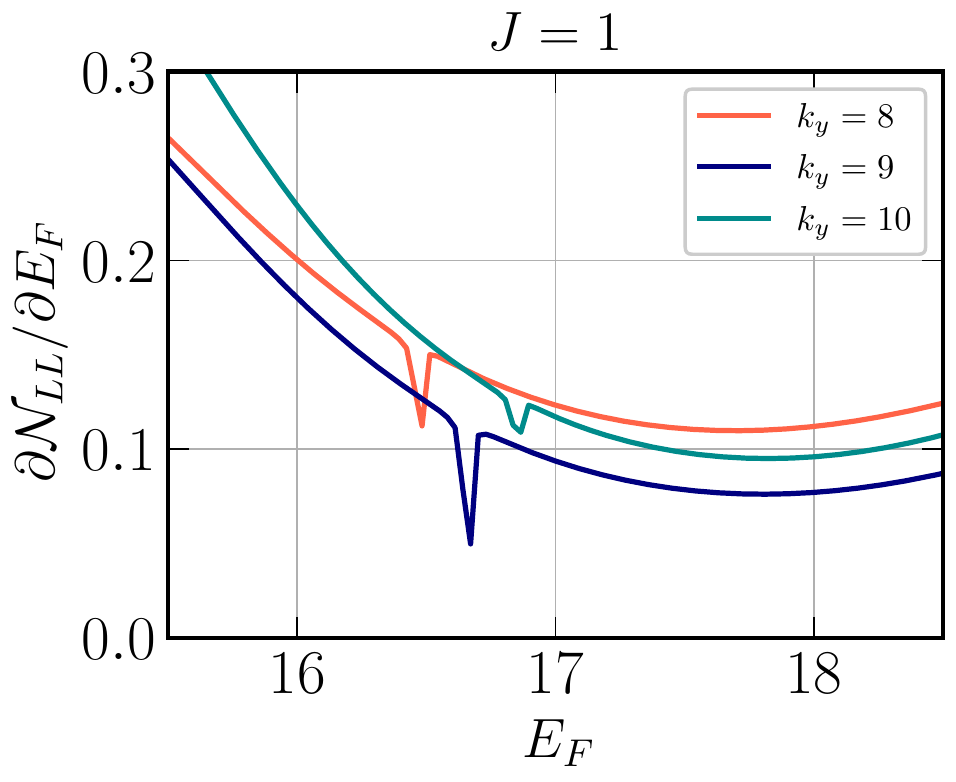}}\\
	\subfigure[]{\includegraphics[width= 0.318\textwidth]{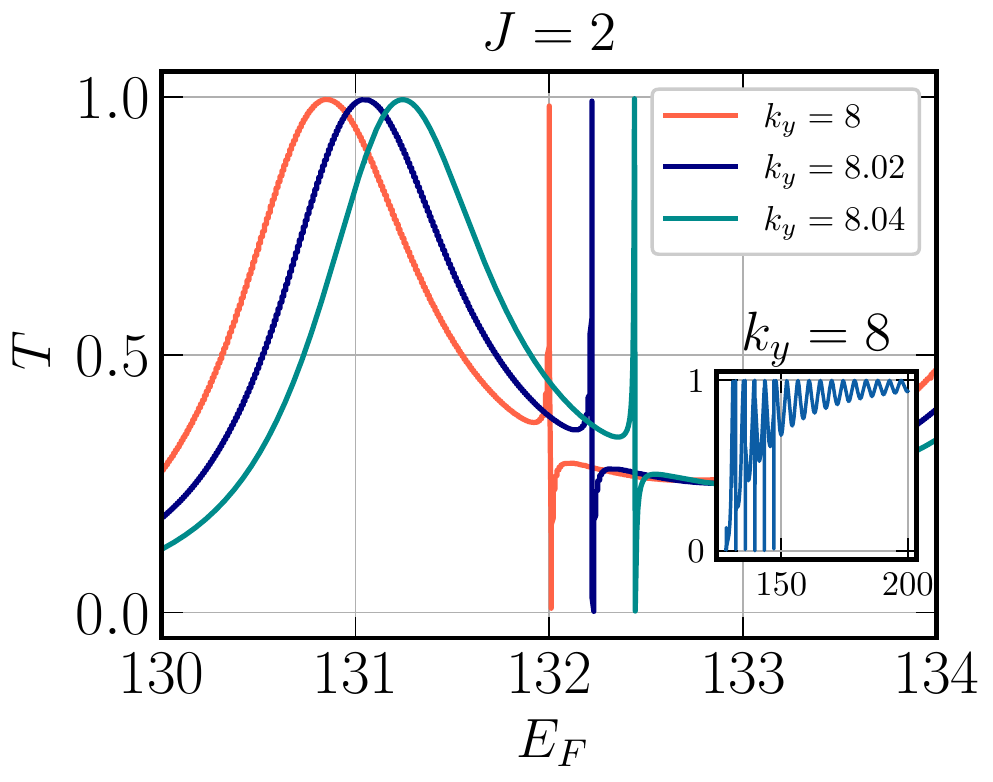}}\quad
	\subfigure[]{\includegraphics[width= 0.308 \textwidth]{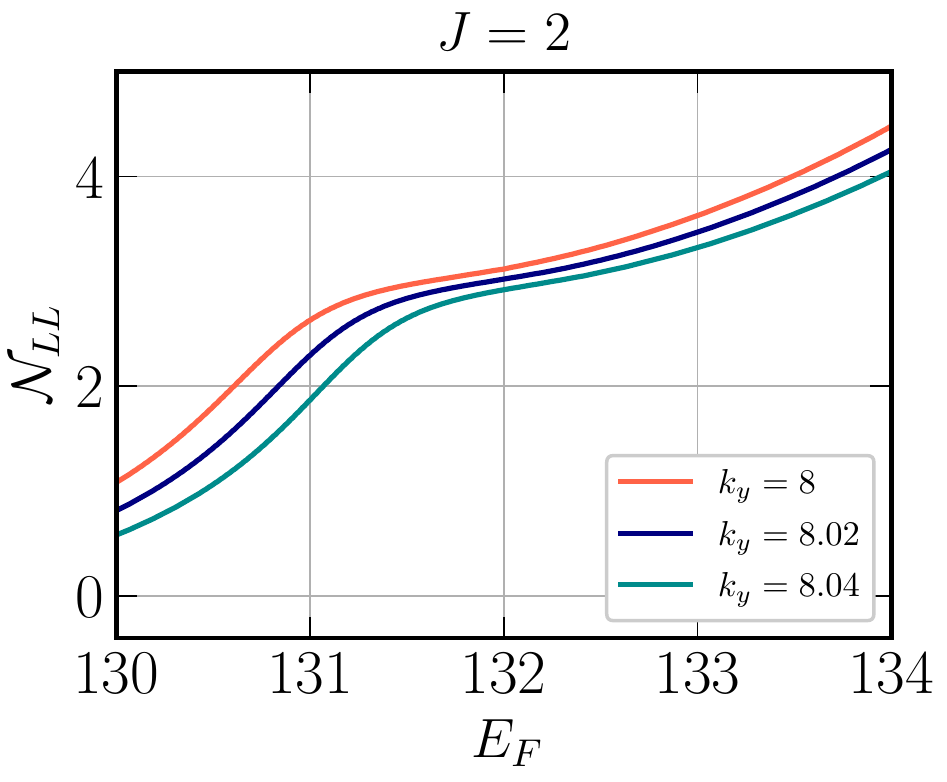}}\quad
	\subfigure[]{\includegraphics[width= 0.338\textwidth]{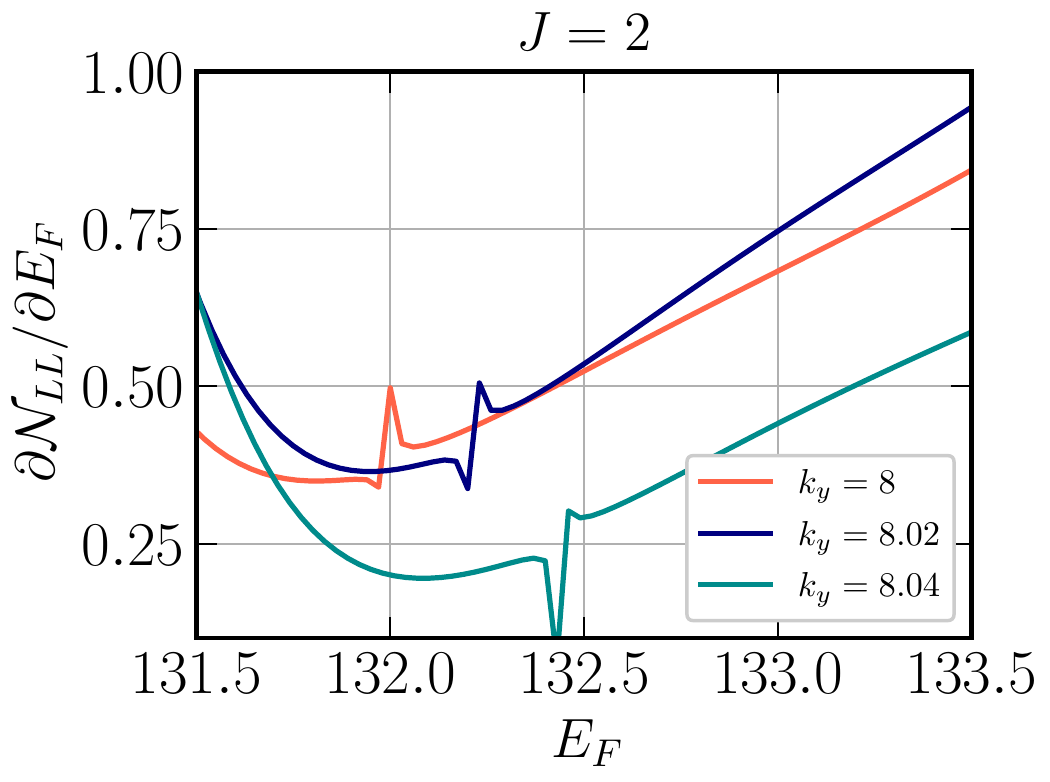}}\\
	\subfigure[]{\includegraphics[width= 0.33 \textwidth]{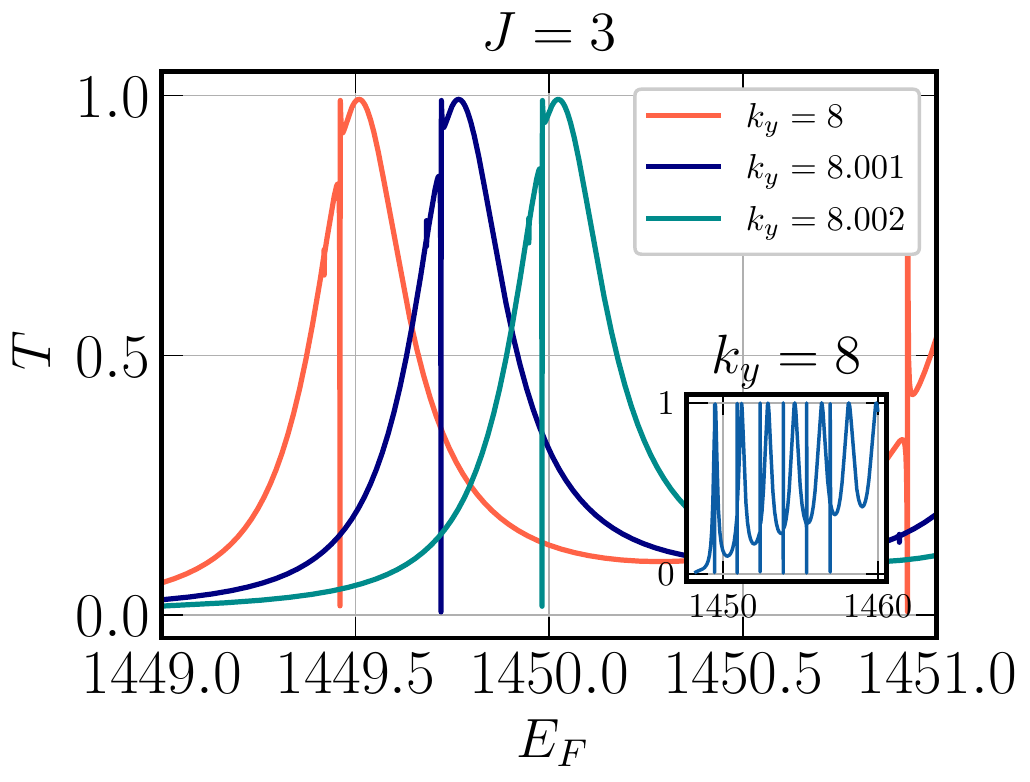}}\quad
	\subfigure[]{\includegraphics[width= 0.32\textwidth]{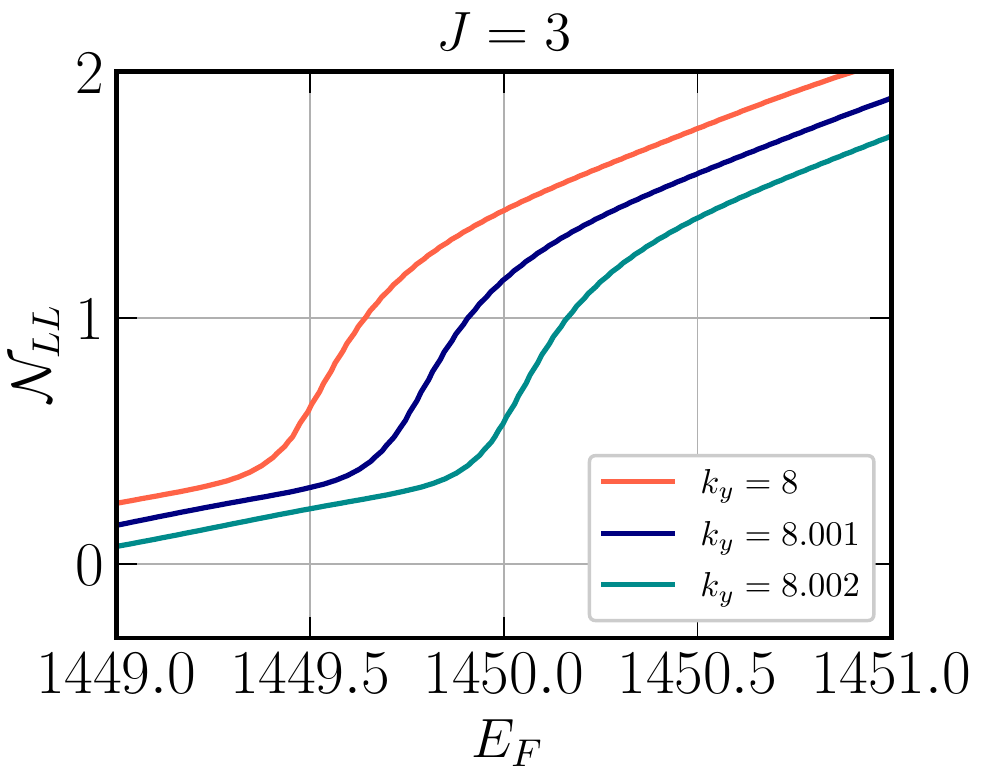}}\quad
	\subfigure[]{\includegraphics[width= 0.315\textwidth]{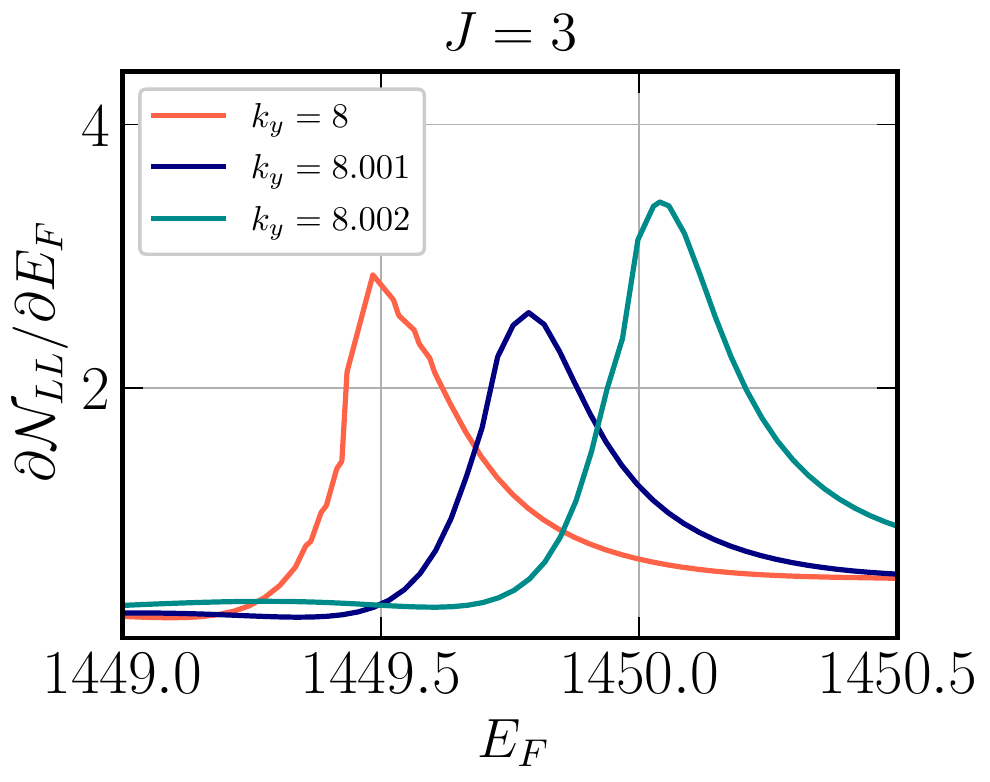}}
	\caption{\label{fig:mweyl_kz}Weyl and multi-Weyl semimetals: Transmission coefficient $T$, pumped shot noise $\mathcal{N}_{LL}$ (in units of $2\,\pi\times 10^{-2}\, v_{\perp} k_0$, remembering that $e= c=\hbar=1$ in the natural units that we have used), and differential pumped shot noise $\partial \mathcal{N}_{LL}/ \partial E_F$ (in units of $2\,\pi\times 10^{-2}$) are shown as functions of the Fermi energy $E_F$ (in units of $v_{\perp}  k_0 $), with the potential well oriented along the $z$-axis. In each plot, three different values of $k_y$ (in units of $k_0$) have been displayed, as indicated in the plot-legends. The values of the remaining parameters used are $V_0=80\, v_{\perp} k_0$, $V_1=2 \, v_{\perp} k_0$, $\hbar\omega=20\, v_{\perp} k_0$, $k_x=8\, k_0$, and $L=0.6\, k_0^{-1}$. The insets in the left-most panels show the behaviour of $T$ for larger intervals of $E_F$, with $ k_y = 8\,k_0$.}
\end{figure}

\subsection{Weyl and multi-Weyl semimetals} 
\label{sec:mweyl}

The low-energy continuum Hamiltonian for a Weyl/multi-Weyl semimetal is given by \cite{liu2017predicted,lundgren2014thermo,fang2012multi,Gang2011,Mukherjee2018}
\begin{align} 
	\label{eq:mweyl_ham}
	\mathcal{H}_{\t{mw}} =   \alpha_J 
	\left	(k_{-}^J \, \sigma_{+} + k_{+}^J \,\sigma_{-} \right ) 
	+ \chi\,v_z \, k_z \, \sigma_z \,  ,
\end{align}
where $\chi=\pm 1$ refers to chirality, and here we set $\chi=1$ without any loss of generality.
The velocities $v_z$ and $v_\perp$ describe the Fermi velocities along the directions of the $z$-axis and perpendicular to it, respectively. Furthermore, $\alpha_J = v_{\perp}/k_0^{J-1} $, where $k_0$ is a material-dependent parameter with the dimensions of momentum, $k_\pm = k_x \pm i \, k_y$, $\sigma_\pm = (\sigma_x \pm i \, \sigma_y)/2$, and $J$ represents the magnitude of the monopole charge at the multi-Weyl node. 
The space-group symmetries restrict $J$ to be less than or equal to three.
Note that $J=1$ represents a Weyl node, in which case $v_z$ is equal to $v_\perp$ (since it is isotropic).
$J=2$ and $J=3$ represent the double-Weyl and triple-Weyl semimetals, respectively.

The two energy bands of the Hamiltonian in Eq.~\eqref{eq:mweyl_ham} are given by
\begin{align}
	\label{eqenweyl}
	\mathcal{E}_{\t{mw}}^\pm(\*k) = \pm  \sqrt{\alpha_J^2 \, k_\perp^{2J} 
		+ v_z^2 \, k_z^2} \, ,
\end{align}
where $k_\perp = \sqrt{k_x^2+k_y^2}$. This is a gapless spectrum with the two energy bands crossing each other at
$\*k=0$. For $J=1$, the dispersion is isotropic and linear in momentum, representing the Weyl semimetals.
Setting $J=2$ or $3$ makes the dispersion quadratic or cubic in the $xy$-plane, while the dispersion remains linear along the $z$-direction. 

\begin{figure}[] 
	\centering
	\subfigure[]{\includegraphics[width= 0.315\textwidth]{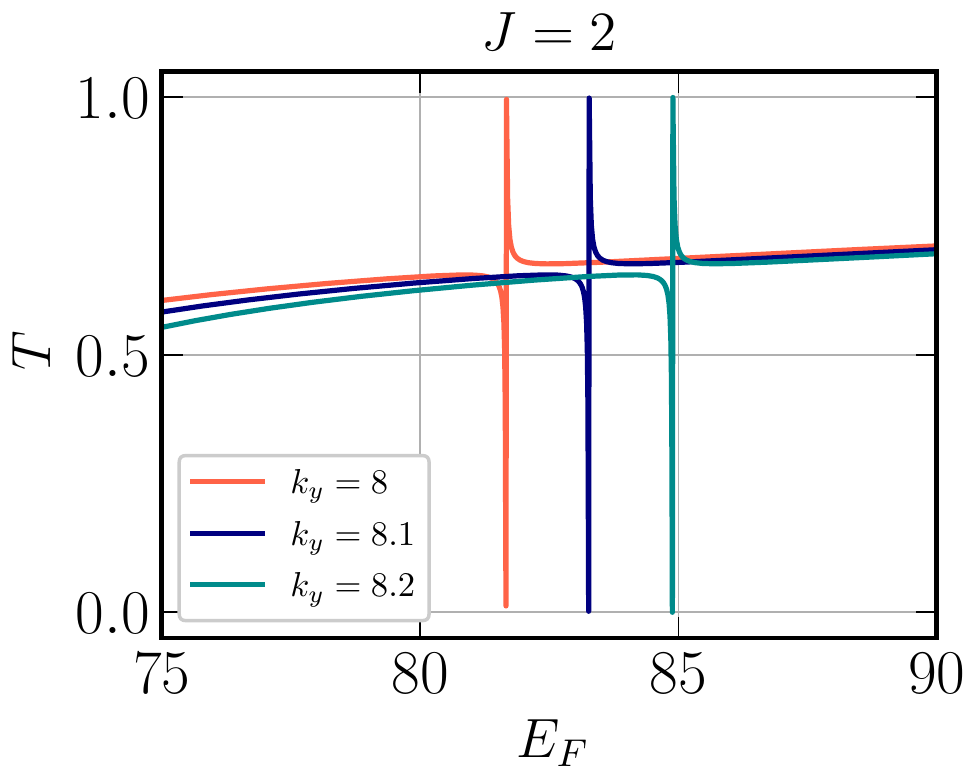}}~
	\subfigure[]{\includegraphics[width= 0.315\textwidth]{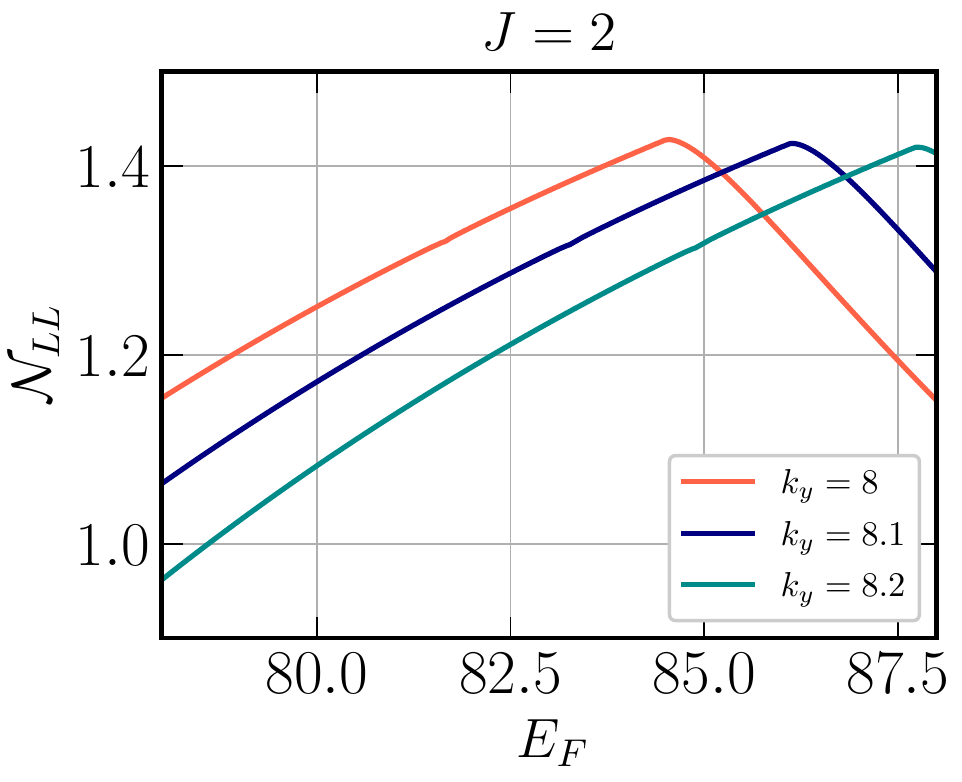}}~
	\subfigure[]{\includegraphics[width= 0.32\textwidth]{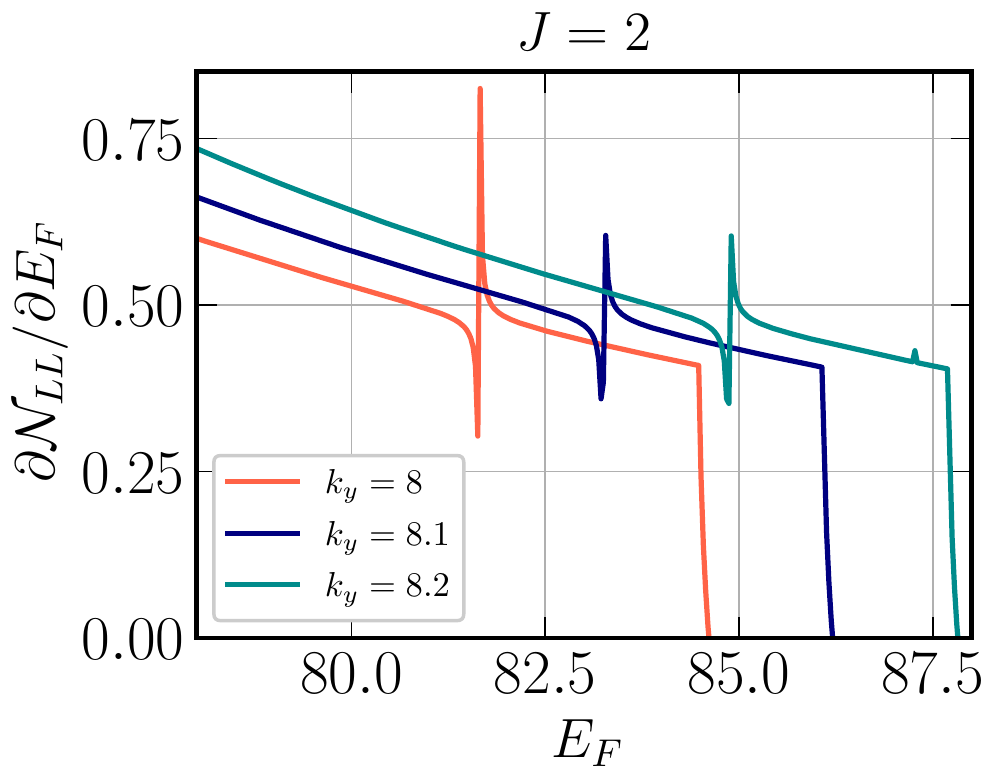}}	\\
	\subfigure[]{\includegraphics[width= 0.315 \textwidth]{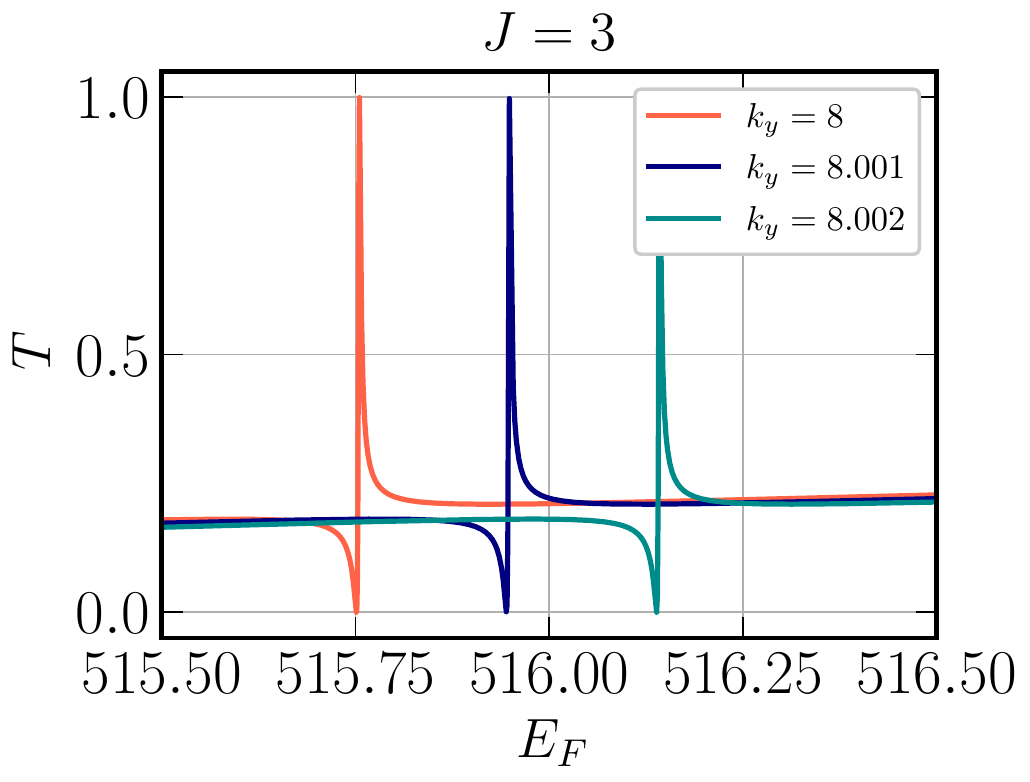} }
	\subfigure[]{\includegraphics[width= 0.33\textwidth]{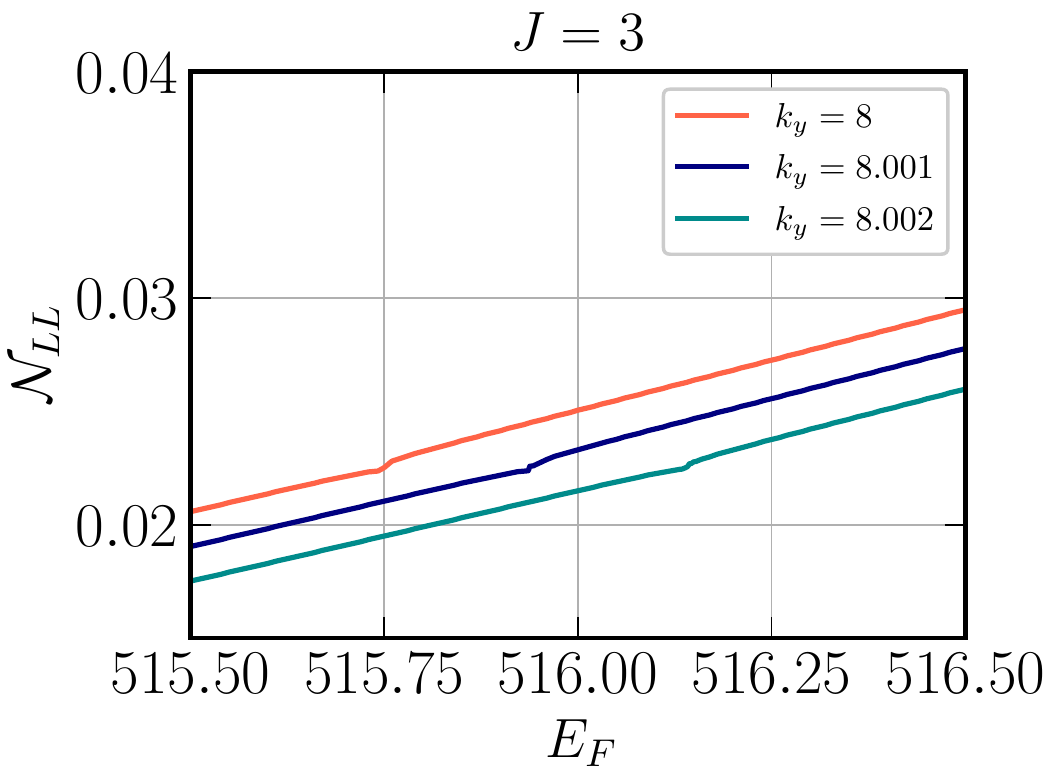}}
	\subfigure[]{\includegraphics[width= 0.31\textwidth]{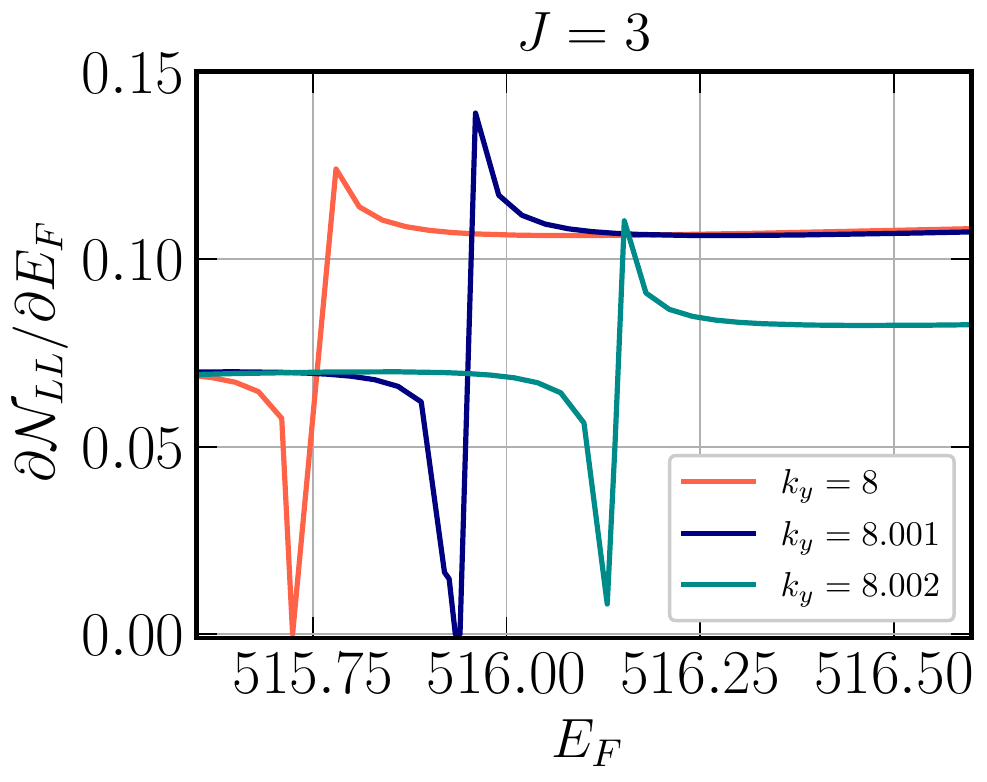}}
\caption{\label{fig:mweyl_kx}Multi-Weyl semimetals: Transmission coefficient $T$, pumped shot noise $\mathcal{N}_{LL}$ (in units of $2\,\pi\times 10^{-2}\, v_{\perp} k_0$, remembering that $e= c=\hbar=1$ in the natural units that we have used), and differential pumped shot noise $\partial \mathcal{N}_{LL}/ \partial E_F$ (in units of $2\,\pi\times 10^{-3} $) are shown as functions of the Fermi energy $E_F$ (in units of $v_{\perp} k_0 $), with the potential well oriented along the $x$-axis. In each plot, three different values of $k_y$ (in units of $k_0$) have been displayed, as indicated in the plot-legends. The values of the remaining parameters used are $V_0=80\, v_{\perp} k_0$, $V_1=2 \, v_{\perp} k_0$, $\hbar\omega=20\, v_{\perp} k_0$, $k_z=8\, k_0$, and $L=0.4\, k_0^{-1}$.}
\end{figure}

In our numerics, we scale our Hamiltonian by $v_{\perp} k_{0}$:
\begin{align}
\frac{\mathcal{H}_{\t{mw}}}{v_{\perp} \,k_{0}} &= 
	\left ( \frac{k_-}{k_0}  \right )^J  \sigma_{+}
	+ \left ( \frac{k_+}{k_0}  \right )^J \sigma_{-}
	+ \chi \,\frac{v_{z} \,k_{z}} {v_{\perp}\, k_0} \sigma_{z}\,,
\end{align}
such that all momentum components are measured in units of $k_0$, energy is measured in units of $ v_{\perp} k_0$ (where we have set $\hbar=1$), and the length scales are in units of $1/k_0$. For calculational simplicity, we set $v_z=v_{\perp}$ for all $J$ values. 

First, we consider a potential well along the $z$-direction. In this case, the functions in Eq.~\eqref{eq:psi}
take the forms:
\begin{align}
	\label{eqfnmwz}
	& {f^{\text{i}}} _{n1} = \frac{\zeta_n+k_n v_r} {n_1 \,k_+   ^J} \,, \quad
	 {f^{\text{i}}} _{n2} = \frac{1}{n_1}, \quad 
	 {f^{\text{o}}} _{n1} = \frac{\zeta_n-k_n v_r} {n_2\,k_+   ^J}\,, \quad 
	 {f^{\text{o}}} _{n2} = \frac{1}{n_2}\,, 
	\nn &
	 {\tilde{f}^{\text{i}}} _{m1} = \frac{\zeta_m+q_m v_r} {n_3\,k_+   ^J}\,, \quad
	 {\tilde{f}^{\text{i}}} _{m2} = \frac{1}{n_3}, \quad 
	 {\tilde{f}^{\text{o}}} _{m1} = \frac{\zeta_m-q_m v_r} {n_4 \,k_+   ^J}\,, \quad
	 {\tilde{f}^{\text{o}}} _{m2} = \frac{1}{n_4}, \nn
	&n_1 = \sqrt{\frac{(\zeta_n+k_n v_r)^2} {k_\perp^{2J}   }+1}\,, \quad 
	n_2 = \sqrt{\frac{(\zeta_n-k_n v_r)^2} {k_\perp^{2J}   }+1}\,, \quad 
	n_3 = \sqrt{\frac{(\zeta_m+q_m v_r)^2} {k_\perp^{2J}   }+1}\,, \quad 
	n_4 = \sqrt{\frac{(\zeta_m-q_m v_r)^2} {k_\perp^{2J}   }+1}\,, \nn 
	&k_n = \frac{1}{v_r} \sqrt{E_n^2-k_\perp^{2J} } \,, \quad 
	q_m = \frac{1}{v_r} \sqrt{(E_m+V_0)^2-k_\perp^{2J}   } \,, \quad
	\zeta_n = \sqrt{v_r^2 \,k_n^2 + k_\perp^{2J}  } \,, \quad
	\zeta_m = \sqrt{v_r^2 \, q_m^2 + k_\perp^{2J} } \,, 
	\quad v_r = \frac{v_z} {v_{\perp}} \, .
\end{align}
As pointed out above, we have set $v_r=1$ for the sake of simplicity.
Appendix~\ref{appA} (in particular, Appendix~\ref{appweylz}) contains more details regarding the computation of the S-matrix.
The transmission coefficient $T$, pumped shot noise, and differential pumped shot noise, as functions of $E_F$, are shown in Fig.~\ref{fig:mweyl_kz}, for some representative parameter values.
Fano resonances are seen for each system, and we denote the corresponding values of $E_F$ by $\efano$. 
For the Weyl semimetal, for example, we find $ \efano =16.48$ for $k_x=k_y=8$ [cf. the orange curve in Fig.~\ref{fig:mweyl_kz}(a)]. Similar to the cases of the 2D electron gas/graphene \cite{Zhu15} and the quadratic band-touching semimetals \cite{Bera2021}, and unlike the 2D pseudospin-1 Dirac-Weyl system \cite{Zhu17}, the Fano resonance curves here are asymmetric, because each curve shows a dip followed by a peak (or vice versa) at $E_F \simeq \efano$.
We note that the resonant energy increases for higher values of $k_y$, similar to a 2D electron gas, but in contrast with graphene \cite{Zhu15}. The characteristics of $T$ for larger intervals of $E_F$ are illustrated in the insets, which show the existence of more Fano resonance points in the spectrum. 
We would like to point out that as $J$ increases, there is a significant increase in $\efano$. The reason for this behaviour can be understood from the discussions in Sec.~\ref{sec:boundstates}, where we illustrate how $\efano$ is related to the bound states of the potential well.
While a Fano resonance is characterized by a quick drop followed by a steep rise in the Weyl ($J=1$) and triple-Weyl ($J=3$) semimetals, the behaviour is just the opposite in the double-Weyl ($J=2$) semimetal. The resonance points are also imprinted in the pumped shot noise curves, as seen in Fig.~\ref{fig:mweyl_kz}, where their signature is captured by inflection points. The differential pumped shot noise profiles, also shown in Fig.~\ref{fig:mweyl_kz}, amplify these signatures.

\begin{figure}[] 
	\centering
	\subfigure[]{\includegraphics[width= 0.32\textwidth]{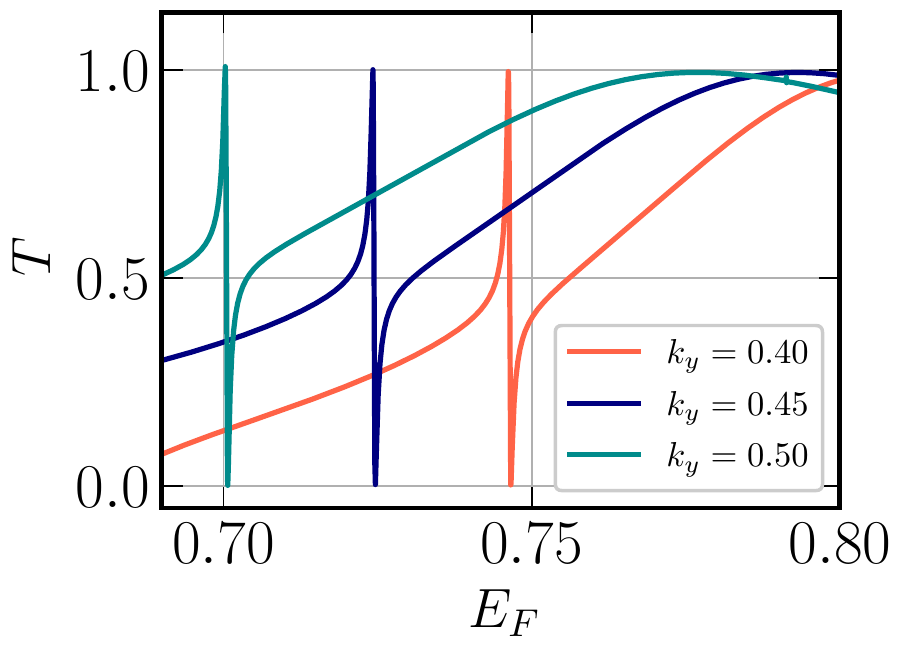}}\quad
	\subfigure[]{\includegraphics[width= 0.313 \textwidth]{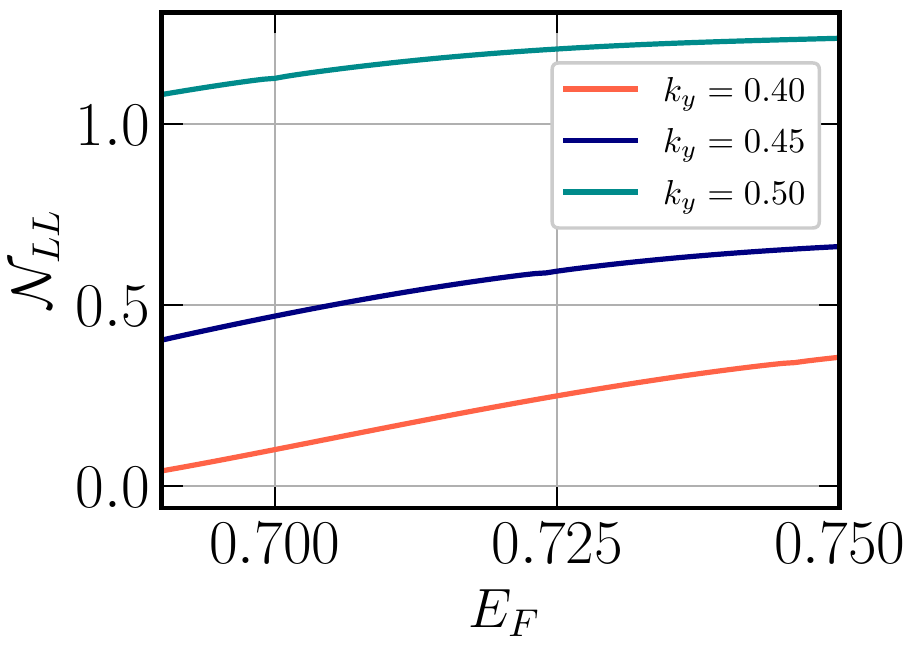}}\quad
	\subfigure[]{\includegraphics[width= 0.315 \textwidth]{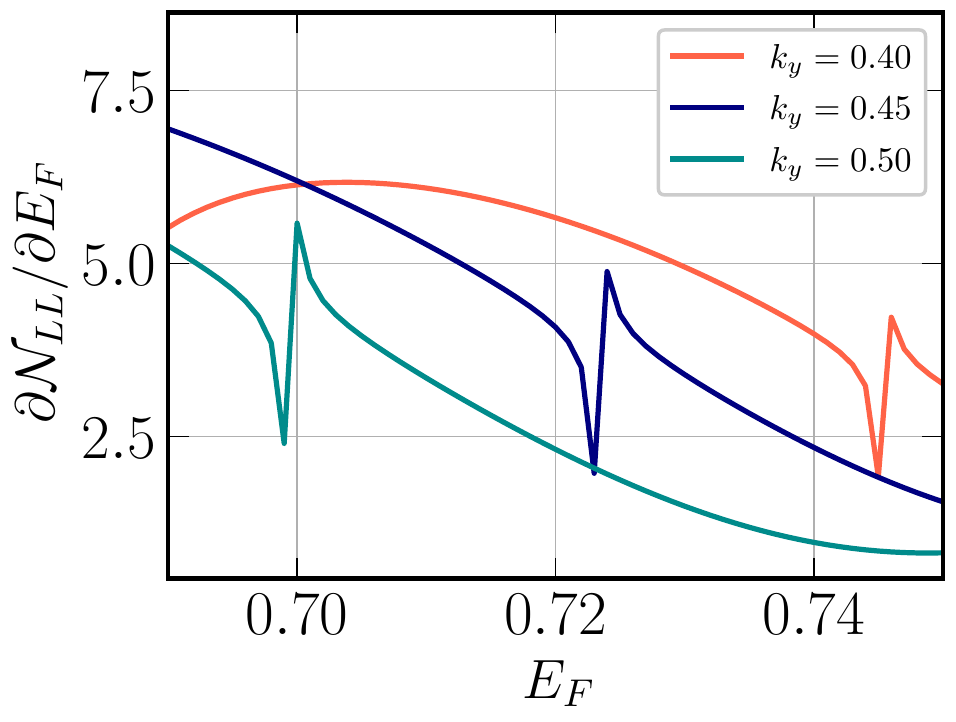}}
	\caption{\label{fig:nodalline_kz}Nodal-line semimetal: (a) Transmission coefficient $T$, (b) pumped shot noise $\mathcal{N}_{LL}$ (in units of $2\,\pi\times 10^{-3}\, {\mathcal M}$), and (c) differential pumped shot noise $\partial \mathcal{N}_{LL}/ \partial E_F$ (in units of $2\,\pi\times 10^{-3} $) are shown as functions of the Fermi energy $E_F$ (in units of ${\mathcal M}$), with the potential well oriented along the $z$-axis. For the units, we remind the reader that $e= c=\hbar=1$ in the natural units that we have used. In each plot, three different values of $k_y$ (in units of ${\mathcal M}$) have been displayed, as indicated in the plot-legends. The values of the remaining parameters used are $V_0= {\mathcal M}$, $V_1= \left( 1/40 \right) {\mathcal M}$, $\hbar\omega= 0.25 \, {\mathcal M} $, $k_x = 0.4\,{\mathcal M} $, $B = {\mathcal M}^{-1}$, and $L= 15 \, {\mathcal M}^{-1} $.}
\end{figure}

Next, we orient the potential well along the $x$-direction, and study the cases for $J>1$, 
\footnote{The $J=1$ case is of course isotropic, and no difference in transport features emerges by changing the orientation of the well.} 
as the dispersions are anisotropic for these systems.
In this case, the functions in Eq.~\eqref{eq:psi}
take the forms:
\begin{align}
	\label{eqfnmwx}
	& {f^{\text{i}}} _{n1} = \frac{(-i)^J (k_y+i \, k_n)^J} {n_1\, (E_n-k_z)}\,, \quad
	 {f^{\text{i}}} _{n2} = \frac{1}{n_1}, \quad 
	 {f^{\text{o}}} _{n1} = \frac{(-i)^J(k_y-i \, k_n)^J} {n_1 \,(E_n-k_z)}\,, \quad 
	 {f^{\text{o}}} _{n2} = \frac{1}{n_1}, \nn
	& {\tilde{f}^{\text{i}}} _{m1} = \frac{(-i)^J(k_y+i \,  q_m)^J} {n_3\,(E_m+V_0-k_z)}\,, \quad
	 {\tilde{f}^{\text{i}}} _{m2} = \frac{1}{n_3}, \quad 
	 {\tilde{f}^{\text{o}}} _{m1} = \frac{(-i)^J(k_y-i \,  q_m)^J} {n_3\,(E_m+V_0-k_z)}\,, \quad
	 {\tilde{f}^{\text{o}}} _{m2} = \frac{1}{n_3}, \nn
	&n_1 = \sqrt{\frac{(k_y^2+k_n^2)^J} {(E_n-k_z)^2}+1}\,, \quad 
	n_3 = \sqrt{\frac{(k_y^2+q_m^2)^J} {(E_m+V_0-k_z)^2}+1}\,, \nn
	&k_n = \sqrt{(E_n^2-k_z^2)^{1/J}-k_y^2}\,, \quad 
	q_m = \sqrt{\left[ (E_m+V_0)^2-k_z^2 \right ] ^{1/J}-k_y^2} \, .
\end{align}
Appendix~\ref{appA} (in particular, Appendix~\ref{appweylx}) contains more details regarding the computation of the S-matrix.
Fig.~\ref{fig:mweyl_kx} shows the characteristics for $T$, pumped shot noise, and differential pumped shot noise, as functions of $E_F$, for some representative parameter values. We notice that the Fano resonances occur at smaller Fermi energies, compared to the case when the well is oriented along the $z$-axis. For example, the first Fano resonance occurs at $\efano=515.7$ for $J=3$, when $k_x= k_y =8$, and this value of $\efano$ is much lower than that seen in Fig.~\ref{fig:mweyl_kz}(c) for similar parameter values. As before, the evidence of the Fano resonances is reflected in the shot noise curves via the inflection points. We would like to point out that while $T$ has more Fano resonance points for a well along the $z$-axis, as we consider larger energy intervals of $E_F$ [cf. the insets of Fig.~\ref{fig:mweyl_kz}], this does not happen in the current situation with a well aligned along the $x$-axis. Hence, for this case, we do not include any insets involving larger energy ranges.

\subsection{Nodal-line semimetals} 
\label{sec:nodalline}

\begin{figure}[] 
	\centering
	\includegraphics[width= \columnwidth]{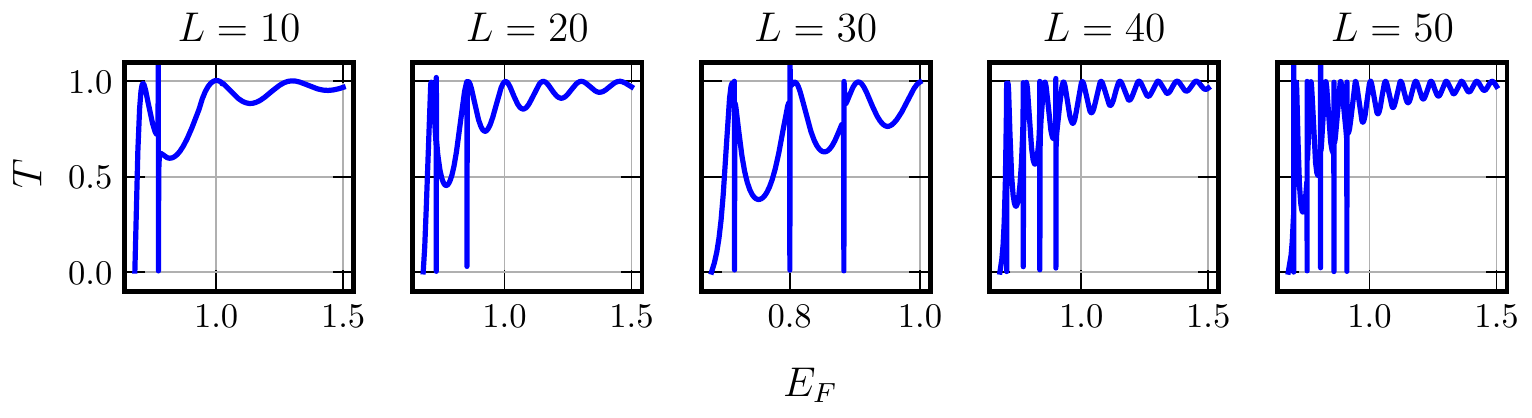}
	\caption{\label{fig:L_variation}
Nodal-line semimetal: The transmission coefficient $T$ is shown, as a function of $E_F$ (in units of ${\mathcal M}$), when the potential well is oriented along the $z$-axis. The series of plots shows the change in behaviour of $T$ as $L$ (in units of ${\mathcal M} ^{-1} $) is increased. The values of the remaining parameters used are $V_0= {\mathcal M}$,
		$V_1= \left( 1/40 \right) {\mathcal M}$,
		$\hbar\omega= 0.25 \, {\mathcal M} $, $k_x =k_y = 0.4\,{\mathcal M}$, and $B = {\mathcal M}^{-1}$.
	}
\end{figure}

In nodal-line semimetals, a band-crossing appears along a closed curve, instead of a single point. We consider a minimal low-energy continuum Hamiltonian \cite{cheng-nodal,Khokhlov2018} for such a system, captured by
\begin{align}
	\label{Hamiltonian_nodal}
	\mathcal{H}_{\t{nl}} = 
	\left	({\mathcal M}- B\, k_{\perp}^{2} \right )  \sigma_{x}
	+  k_z \, \sigma_{z} \, ,
\end{align}
where the nodal-line is a circle $k_x^2+k_y^2= {\mathcal M} /B$, that lies in the $k_z=0$ plane.
The energy bands are given by
\begin{align}
	\label{energy_nodal}
	\mathcal{E}^{\pm}_{\t{nl}} (\*k)=\pm\sqrt{ 
		\left ({\mathcal M}- B\, k_{\perp}^{2} \right )^2+k_z^2}\,.
\end{align}
In our numerics, we scale our Hamiltonian by $\mathcal M$:
\begin{align}
	\frac{\mathcal{H}_{\t{nl}}} {\mathcal M}
	= \left[ 1-  B\,{\mathcal M}\,\left( \frac{k_{\perp}} {\mathcal M} \right)^2
	\right ]  \sigma_{x}
	+ \frac{ k_z} {\mathcal M} \, \sigma_{z}\,,
\end{align}
such that the energy and all the momentum components are measured in units of $\mathcal M$ (where we have set $\hbar=1$), and $B$ and the length scales are in units of $1/\mathcal M$.

Similar to the Weyl/multi-Weyl case, we first set the potential well along the $z$-axis, for which
the functions in Eq.~\eqref{eq:psi} take the forms:
\begin{align}
	\label{eqnlzf}
	& {f^{\text{i}}} _{n1} = \frac{E_n+k_n}{n_1 \left (1-B \,k^2_{\perp} \right )}\,, \quad
	 {f^{\text{i}}} _{n2} = \frac{1}{n_1}, \quad
	{ {f^{\text{o}}} _{n1} = \frac{E_n-k_n} {n_2 \left (1-B\, k_{\perp}^2 \right )}}\,, \quad
	 {f^{\text{o}}} _{n2} = \frac{1}{n_2}, \nn
	& {\tilde{f}^{\text{i}}} _{m1} = \frac{E_m+V_0+q_m}{n_3 \left (1-B\, k_{\perp}^2 \right )}\,, \quad
	 {\tilde{f}^{\text{i}}} _{m2} = \frac{1}{n_3}, \quad
	 {\tilde{f}^{\text{o}}} _{m1} = {\frac{E_m+V_0-q_m}{n_4 \left (1-B\, k_{\perp}^2 \right )}}\,, \quad
	 {\tilde{f}^{\text{o}}} _{m2} = \frac{1}{n_4}, \nn
	&n_1 = \sqrt{\frac{(E_n+k_n)^2} {\left (1-B\, k_{\perp}^2 \right )^2}+1}\,, \quad
	n_2 = \sqrt{\frac{(E_n-k_n)^2} {\left (1-B\, k_{\perp}^2 \right )^2}+1}\,, \quad
	n_3 = \sqrt{\frac{(E_m+V_0+q_m)^2}  {\left (1-B\, k_{\perp}^2 \right )^2}+1}\,, \quad
	n_4 = \sqrt{\frac{(E_m+V_0-q_m)^2}
		{ \left (1-B\, k_{\perp}^2 \right )^2}+1}\,, \nn
	&k_n = \sqrt{E_n^2-(1-B  \,k^2_{\perp})^2}\,, \quad
	q_m = \sqrt{(E_m+V_0)^2-\left (1-B\, k_{\perp}^2 \right )^2} \, .
\end{align}
Appendix~\ref{appA} (in particular, Appendix~\ref{appnodalz}) contains more details regarding the computation of the S-matrix.
The characteristics for $T$, pumped shot noise, and differential shot noise are illustrated in Fig.~\ref{fig:nodalline_kz}, as functions of $E_F$, for some representative parameters. $T$ in Fig.~\ref{fig:nodalline_kz}(a) shows several asymmetric Fano resonance points, one of them being at $\efano=0.746$ for $k_x=k_y=0.4$. Each resonance point features a peak followed by a dip, which is opposite in behaviour to that in the Weyl and triple-Weyl semimetals (because in Figs.~\ref{fig:mweyl_kz}(a) and \ref{fig:mweyl_kz}(c), a dip is followed by a peak), and similar to that observed for a double-Weyl semimetal [cf. Fig.~\ref{fig:mweyl_kz}(b)]. For larger Fermi energies, we find no further resonances. The pumped shot noise is shown in Fig.~\ref{fig:nodalline_kz}(b). Since the inflection points of the pumped shot noise are not very prominent in this case, differentiating the curve with respect to $E_F$ magnifies/highlights the existing inflection points. In this way, the Fano resonance points can be easily identified from the sharp resonances in the differential shot noise, as seen in Fig.~\ref{fig:nodalline_kz}(c). 

Interestingly, we find that for every ten units increment in $L$, the number of the Fano resonance points increases by unity, which is captured in Fig.~\ref{fig:L_variation}.
This can be understood as follows: In Appendix~\ref{appA}, we have shown that the scattering matrix depends on ${\bf M}_{AA}$,  ${\bf M}_{AB}$, ${\bf M}_{BA}$, and ${\bf M}_{BB}$, where the dependence on the system size $L$ is embedded in the exponential terms like $e^{-i \,q_{m} L}$ and $e^{-i\, k_{m} L}$ (or a product of both). Therefore, if the $L$ dependence comes from $e^{-i \,k_m L}$, then the Fano resonance point at $k_m=k^{\text{Fano}}_m$ has to satisfy $e^{-i \,k^{\t{Fano}}_m L}= const $, where $const $ is a real or complex number. This shows that the separation between consecutive Fano resonance points decreases if $L$ is increased. In other words, a larger $L$ facilitates the accommodation of more Fano resonance points within a given energy window.

\begin{figure}[] 
	\centering
	\subfigure[]{\includegraphics[width= 0.31\textwidth]{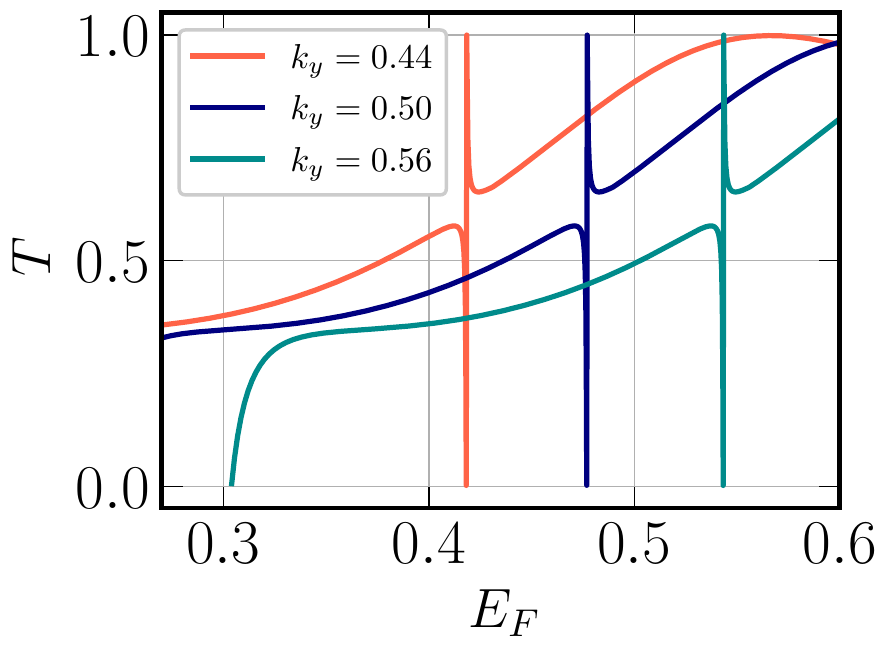}}\quad
	\subfigure[]{\includegraphics[width= 0.304\textwidth]{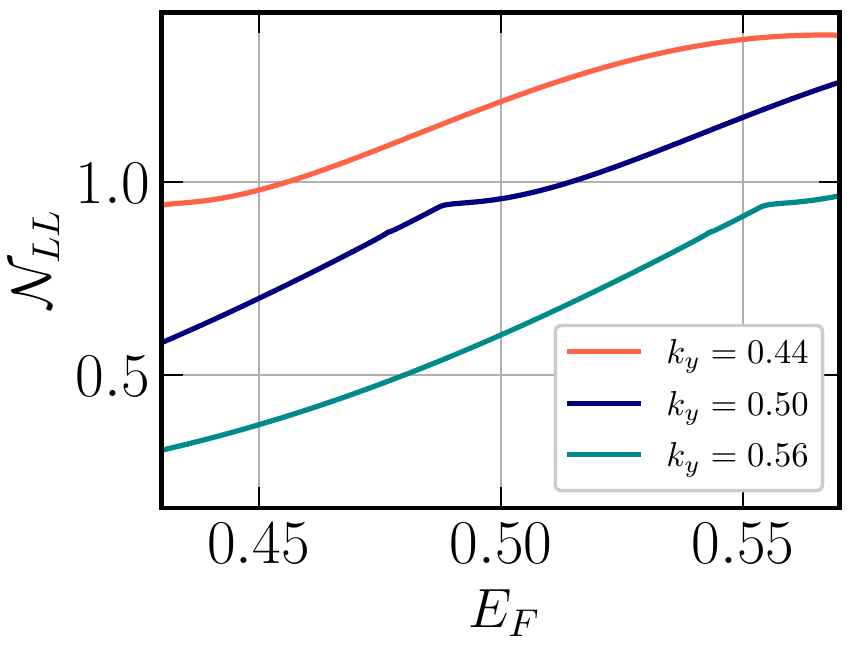}}\quad
	\subfigure[]{\includegraphics[width= 0.305\textwidth]{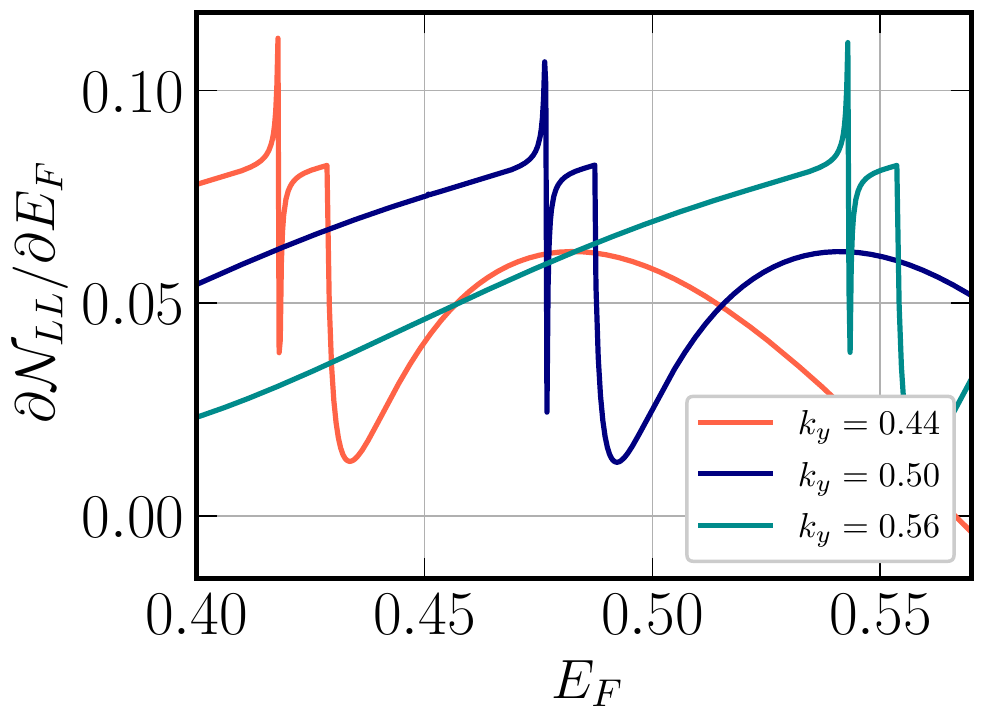}}
	\caption{\label{fig:nodalline_kx}Nodal-line semimetal: (a) Transmission coefficient $T$, (b) pumped shot noise $\mathcal{N}_{LL}$ (in units of $2\,\pi\times 10^{-3}\, {\mathcal M}$), and (c) differential pumped shot noise $\partial \mathcal{N}_{LL}/ \partial E_F$ (in units of $2\,\pi\times 10^{-3}$) are shown as functions of the Fermi energy $E_F$ (in units of ${\mathcal M}$), with the potential well oriented along the $x$-axis. For the units, we remind the reader that $e= c=\hbar=1$ in the natural units that we have used. In each plot, three different values of $k_y$ (in units of ${\mathcal M}$) have been displayed, as indicated in the plot-legends. The values of the remaining parameters used are $V_0= {\mathcal M}$, $V_1= \left( 1/40 \right) {\mathcal M}$, $\hbar\omega= 0.25 \, {\mathcal M} $, $k_z= 0$, $B = {\mathcal M}^{-1}$, and $L= 15 \, {\mathcal M}^{-1} $.}
\end{figure}

Next, we set the potential well along the $x$-axis, for which
the functions in Eq.~\eqref{eq:psi} take the forms:
\begin{align}
	\label{eqnodalx}
	& {f^{\text{i}}} _{n1} = \frac{1-B \,(k_y^2+k_n^2)} {n_1\,(E_n-k_z)}\,, \quad
	 {f^{\text{i}}} _{n2} = \frac{1}{n_1}\,, \quad
	 {\tilde{f}^{\text{i}}} _{m1} = \frac{1-B\, (k_y^2+q_m^2)} {n_2\,(E_m+V_0-k_z)}\,, \quad
	 {\tilde{f}^{\text{i}}} _{m2} = \frac{1}{n_2} \,, \nn
	&   {f^{\text{o}}} _{n1} =  {f^{\text{i}}} _{n1} \,, \quad 
	 {f^{\text{o}}} _{n2} =  {f^{\text{i}}} _{n2} \,, \quad
	 {\tilde{f}^{\text{o}}} _{m1} =  {\tilde{f}^{\text{i}}} _{m1} \,, \quad
	 {\tilde{f}^{\text{o}}} _{m2} =  {\tilde{f}^{\text{i}}} _{m2} \,, \nn
	& n_1 = \sqrt{\frac{\lsb 1-B\, (k_y^2+k_n^2)\rsb^2} {(E_n-k_z)^2}+1}, \quad
	n_2 = \sqrt{\frac{\lsb 1-B\, (k_y^2+q_m^2)\rsb^2} {(E_m+V_0-k_z)^2}+1}, \nn
	& k_n = \sqrt{\frac{1-B\, k_y^2+\sqrt{E_n^2-k_z^2}}{B}}, \quad
	q_m = \sqrt{\frac{1-B\, k_y^2+\sqrt{(E_m+V_0)^2-k_z^2}}{B}} \, .
\end{align}
The derivation of the form of the S-matrix for this case is elaborated on in Appendix~\ref{appnlx}.
As illustrated in Fig.~\ref{fig:nodalline_kx}, $T$ contains Fano resonance points for which a dip is followed by a peak (similar to the multi-Weyl cases illustrated in Fig.~\ref{fig:mweyl_kx}). A representative point exists at $ \efano=0.42$, for $k_y=0.44$ and $ k_z=0$. These resonance points, as before, are also encoded as inflection points in the shot noise, which are magnified in the differential shot noise spectrum. We would like to point out that unlike the case of the potential aligned along the $z$-axis, the number of resonance points does not increase with $L$ for this case.

\section{Quasi-bound states of the static well} 
\label{sec:boundstates}

In this section, we investigate the confined states $E_b$ of the static quantum well of length $L$ and fixed depth $V_0$. These are obtained by setting $m=n=0$ in Eq.~\eqref{eq:psi}. When we can solve for the coefficients by implementing the continuity of the wavefunction only, we get the secular equation
\begin{align}
	\begin{vmatrix}
		 {f^{\text{i}}} _{01}\, e^{-\rho} & - {\tilde{f}^{\text{i}}} _{01} \,e^{-\sigma} & - {\tilde{f}^{\text{o}}} _{01} \,e^{\sigma} & 0 \\ 
		0 &  {\tilde{f}^{\text{i}}} _{01}\, e^{\sigma} &  {\tilde{f}^{\text{o}}} _{01}\, e^{-\sigma} & - {f^{\text{o}}} _{01} \,e^{-\rho} \\
		 {f^{\text{i}}} _{02} \,e^{-\rho} & - {\tilde{f}^{\text{i}}} _{02} \,e^{-\sigma} & - {\tilde{f}^{\text{o}}} _{02}\, e^{\sigma} & 0 \\
		0 & - {\tilde{f}^{\text{i}}} _{02} \,e^{\sigma} & - {\tilde{f}^{\text{o}}} _{02} \,e^{-\sigma} & { {f^{\text{o}}} _{02} \,e^{-\rho}}
	\end{vmatrix}
	= 0\,,
	\label{eqroots1}	
\end{align}
where $\rho=k_0\, L/2$ and $\sigma=i \,q_0 L/2$. 
The roots of this transcendental equation then give the energies $E_b$ of the bound states of the static well, 
which have to be determined numerically. 

\subsection{Weyl/multi-Weyl semimetal with potential well aligned along the $z$-axis}
\label{boundweylz}

\begin{figure}[] 
	\centering
	\subfigure[]{\includegraphics[width= 0.32\textwidth]{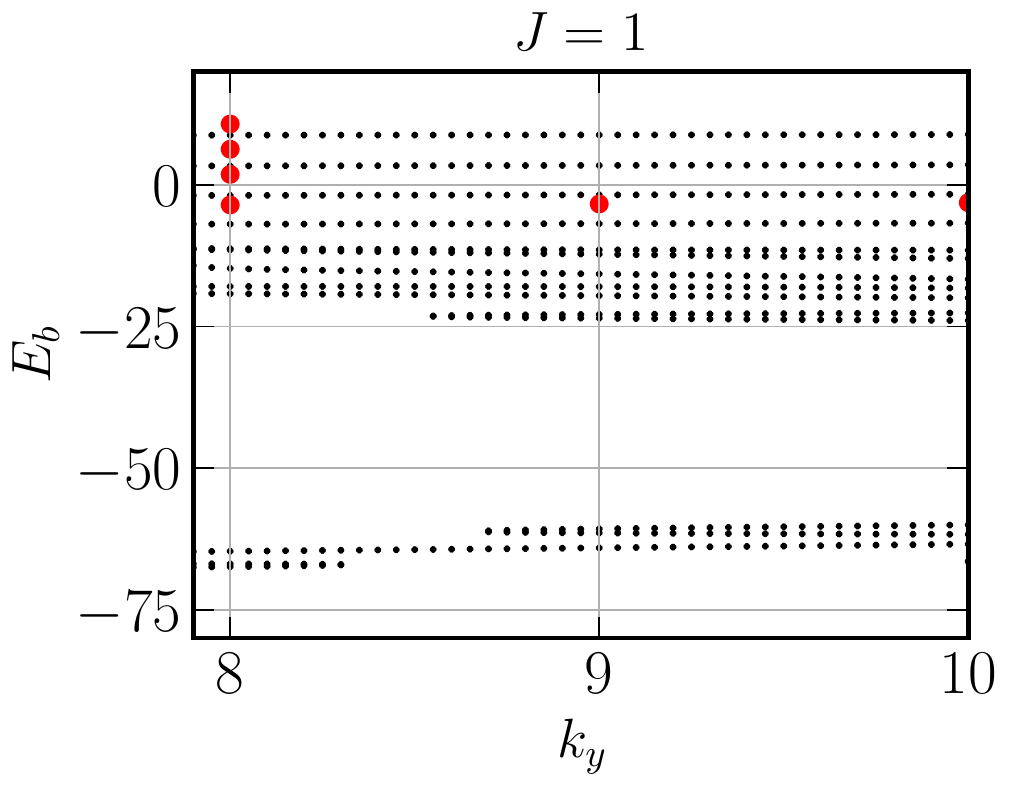}}\quad
	\subfigure[]{\includegraphics[width= 0.32\textwidth]{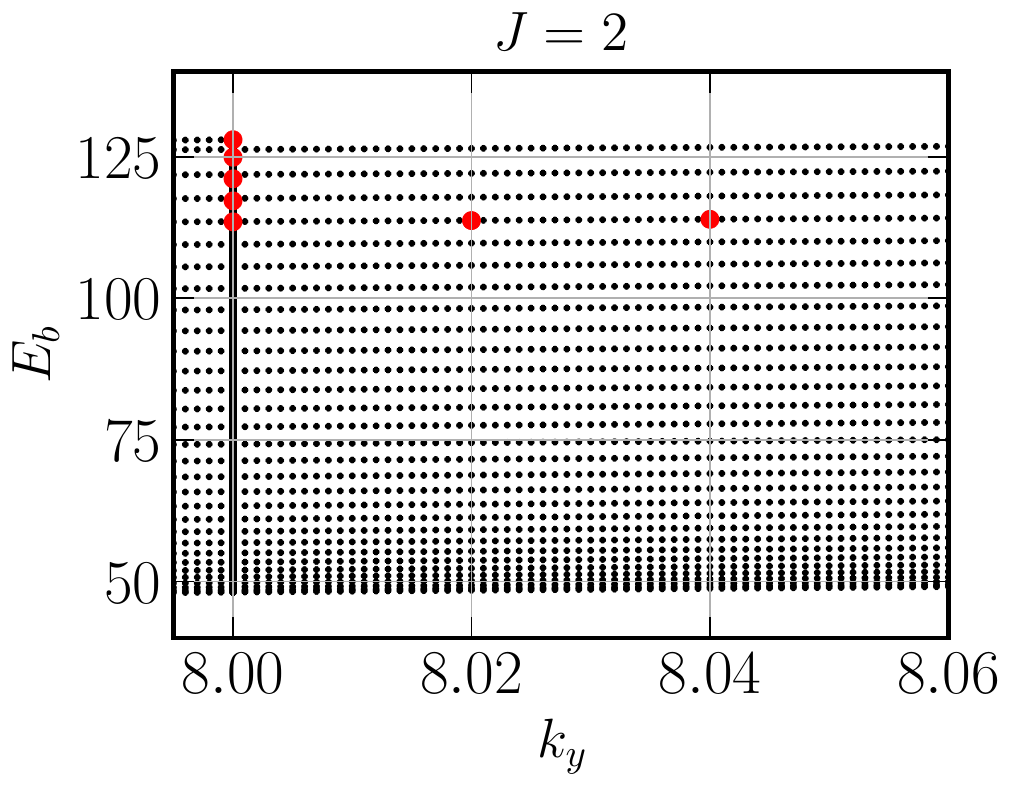}}\quad
	\subfigure[]{\includegraphics[width= 0.32\textwidth]{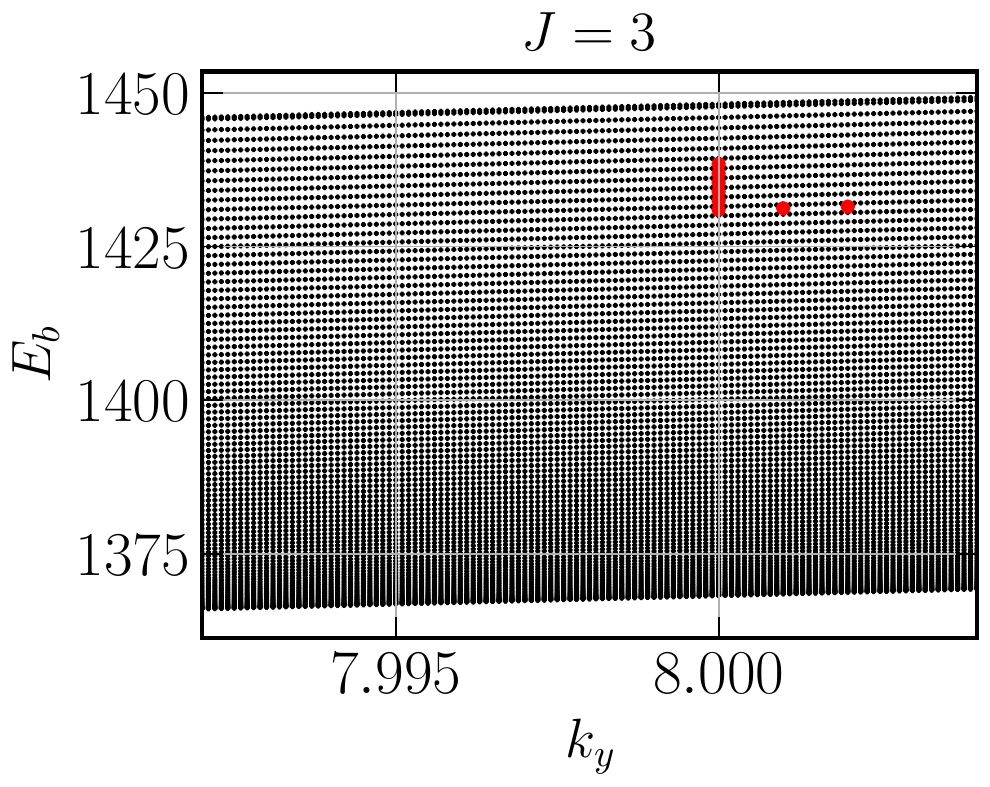}}
	\caption{\label{fig:mweyl_boundstate_kz}Weyl and multi-Weyl semimetals: The bound states of a static quantum well, oriented along the $z$-direction, are shown in black dots for (a) $J=1$, (b) $J=2$, and (c) $J=3$. The red dots represent the values of $E_b$, which coincide with the energies of the Fano resonance points in Fig.~\ref{fig:mweyl_kz} for the driven well (with $V_1 = 2 \, v_{\perp} k_0$ and $\hbar\omega=20\, v_{\perp} k_0$), such that $E_b = \efano -\hbar \omega$. The values of the remaining parameters used are $V_0=80\, v_{\perp} k_0$, $k_x=8\, k_0$, and $L=0.6\, k_0^{-1}$.}
\end{figure}

For the Weyl and multi-Weyl semimetals, with the well aligned along the $z$-axis, Eq.~\eqref{eqroots1} is applicable, after the expressions from Eq.~\eqref{eqfnmwz} have been fed in. The black points in Fig.~\ref{fig:mweyl_boundstate_kz} illustrate the $E_b$-values as functions of $k_y$, using the same parameter values as in Fig.~\ref{fig:mweyl_kz}.
We have earlier emphasized that a Fano resonance is observed when the energy of a bound state matches with that of a Floquet side-band for the incident channel, i.e., $ E_b = \efano -n \hbar \omega $. In the parameter ranges that we consider, this happens when the side-band is of the first order, i.e., when $E_b = \efano-\hbar\omega$ is satisfied. Such points are indicated by red color. The results show, for example, that the Weyl semimetal has four bound states (indicated in red dots) at $k_x=k_y=8$, for $V_0 =80$, which correspond to the Fano resonance points shown in the inset of Fig.~\ref{fig:mweyl_kz}(a). In particular, we have $\efano=16.48$ as one of the resonant energies (with $\hbar\omega=20$), which indeed gives the bound state with $E_b=-3.25$. The same correspondence can be proved for the other red dots in all the plots.

All these bound states satisfy $ \alpha_J \,k_\perp^{J}-V_0 \leq  E_b \leq \alpha_J \,k_\perp^{J} $, as is evident from Eq.~\eqref{eqenweyl}.
This explains why $E_b$ values increase as $J$ is increased. As a result, $\efano $ also increases with $J$ in Fig.~\ref{fig:mweyl_kz}.
Furthermore, our results indicate that the density of the bound states amplifies drastically with an increment in $J$. 

If we increase the driving frequency $\omega$, the side-band energy interval (given by $\hbar \omega$) also increases --- this leads to bound states with higher values of $E_b$ being activated to produce Fano resonances in the transmission spectrum. In other words, a higher value of $\hbar \omega $ increases the possibility of a deeper (quasi)bound state to be activated into the transport process.

\subsection{Multi-Weyl semimetal with potential well aligned along the $x$-axis}
\label{boundweylx}

\begin{figure}[] 
	\centering
	\subfigure[]{\includegraphics[width= 0.32\textwidth]{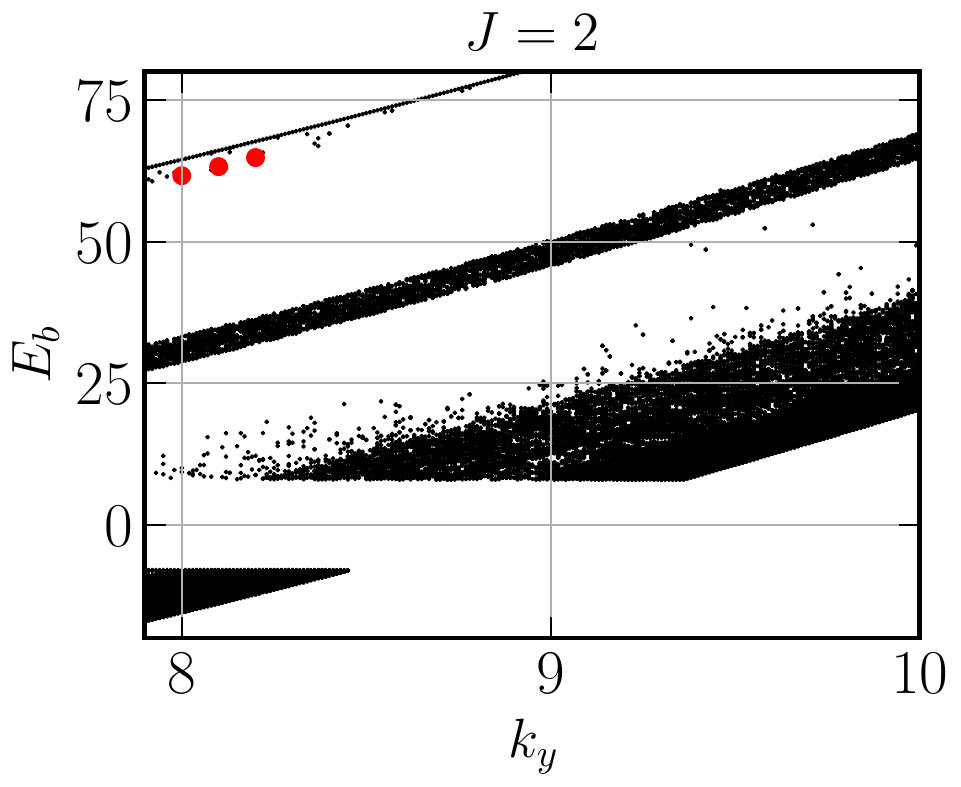}}
	\hspace{2 cm}	
	\subfigure[]{\includegraphics[width= 0.345 \textwidth]{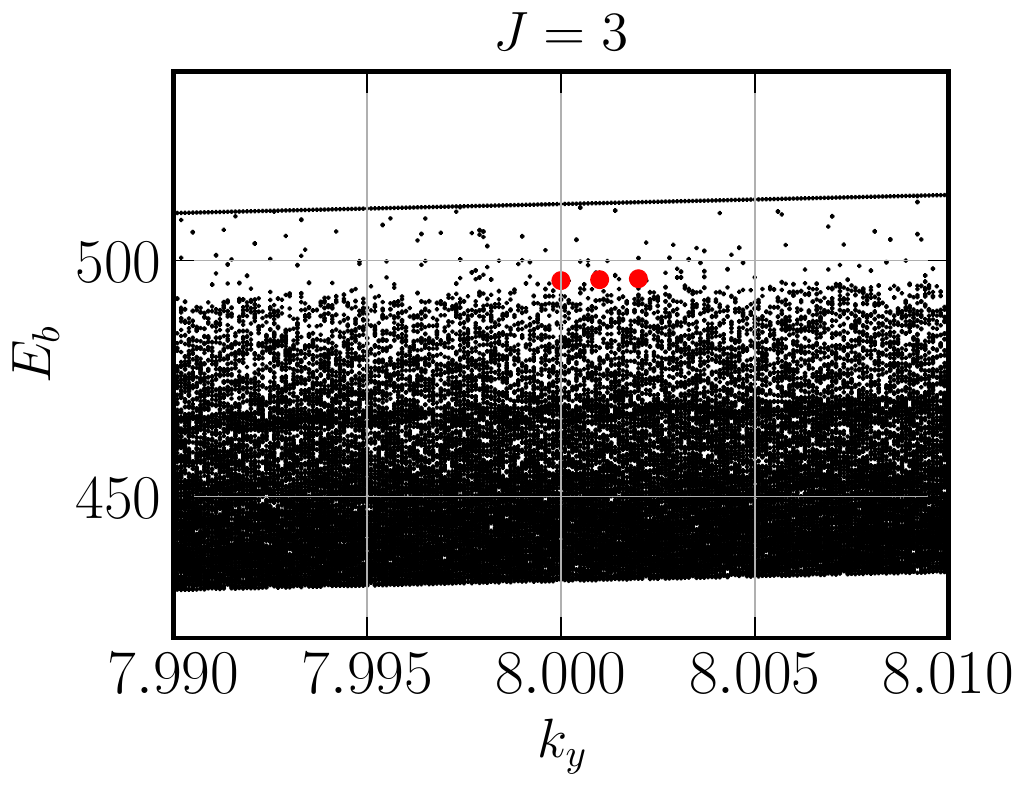}}
	\caption{\label{fig:mweyl_boundstate_kx}
Multi-Weyl semimetals: The bound states of a static quantum well, oriented along the $x$-direction, are shown in black dots for (a) $J=2$ and (b) $J=3$. The red dots represent the values of $E_b$, which coincide with the energies of the Fano resonance points in Fig.~\ref{fig:mweyl_kx} for the driven well (with $V_1 = 2 \, v_{\perp} k_0$ and $\hbar\omega=20\, v_{\perp} k_0$), such that $E_b = \efano -\hbar \omega$.
The values of the remaining parameters used are $V_0=80\, v_{\perp} k_0$, $k_z=8\, k_0$, and $L=0.4 \, k_0^{-1}$.}
\end{figure}

For the multi-Weyl semimetals, anisotropic dispersion implies that the distribution of bound states should change as we align the well along the $x$-axis. We analyze the bound states implementing Eq.~\eqref{eqroots1}, using the expressions from Eq.~\eqref{eqfnmwx}. With the same parameter values as in Fig.~\ref{fig:mweyl_kx}, the black dots in Fig.~\ref{fig:mweyl_boundstate_kx} capture the bound state energies, while the red dots therein highlight the values of $\efano -\hbar \omega$ obtained from the Fano resonances in Fig.~\ref{fig:mweyl_kx}. The plots show large and irregular gaps between the bound states.

For this case, the inequality $ \sqrt{ \alpha_J^2 \,k_y^{2J}+ v_z^2 \,k_z^2 } 
-V_0 \leq E_b \leq \sqrt{ \alpha_J^2 \,k_y^{2J}+ v_z^2 \,k_z^2 } $ is satisfied, as is evident from Eq.~\eqref{eqenweyl}. Hence, the values of $E_b$ are lower than those for the case with the well aligned along the $z$-axis, which results in the Fano resonances showing up at lower energy values in Fig.~\ref{fig:mweyl_kx} (compared to those in Fig.~\ref{fig:mweyl_kz}). The trend of increase in $\efano$,
with an increase in $J$, is also explained by the form of the above inequality.

\subsection{Nodal-line semimetal with potential well aligned along the $z$-axis}
\label{boundnlz}

In case of a nodal-line semimetal, with the potential well aligned along the $z$-axis, the secular equation takes the same form as Eq.~\eqref{eqroots1}, with the expressions of the functions given by Eq.~\eqref{eqnlzf}.
The numerical roots, for the same parameter values as in Fig.~\ref{fig:nodalline_kz}, are shown in Fig.~\ref{fig:nodalline_boundstate}(a), represented by black dots. Here, the bound state energies are bounded by the inequality $ | {\mathcal M}-B\,k_\perp^2 |
-V_0 \le E_b \le |{\mathcal M}-B\,k_\perp^2 |$, as is evident from Eq.~\eqref{energy_nodal}. 

The Fano resonances of Fig.~\ref{fig:nodalline_kz}(a) correspond to the red dots in Fig.~\ref{fig:nodalline_boundstate}(a), obeying $\efano-\hbar\omega = E_b$.
For example, the $T$ in Fig.~\ref{fig:nodalline_kz}(a) shows a resonance at $\efano=0.70$ for $k_x=k_y=0.4$. This corresponds to the bound state with $E_b=0.45$, visible in Fig.~\ref{fig:nodalline_boundstate}(a).

\subsection{Nodal-line semimetal with potential well aligned along the $x$-axis}
\label{boundnlx}

For the nodal-line semimetal, when the potential well is oriented along the $x$-axis, we need to consider both the continuity of the wavefunction and its derivative. This leads to the secular equation
\begin{align}
	\begin{vmatrix}
		 {f^{\text{i}}} _{02} \,e^{-\rho} & - {\tilde{f}^{\text{i}}} _{02} \,e^{-\sigma} & - {\tilde{f}^{\text{i}}} _{02}\, e^{\sigma} & 0 \\ 
		0 &  {\tilde{f}^{\text{i}}} _{02} \,e^{\sigma} &  {\tilde{f}^{\text{i}}} _{02} \,e^{-\sigma} & - {f^{\text{i}}} _{02}\, e^{-\rho} \\
		 {f^{\text{i}}} _{02} \,e^{-\rho} & -\frac{ i \, q_0 }{k_0}  \,  {\tilde{f}^{\text{i}}} _{02} \,e^{-\sigma} 
		& \frac{ i \, q_0 }{k_0} \,  {\tilde{f}^{\text{i}}} _{02}\, e^{\sigma} & 0 \\
		0 & \frac{i \, q_0}{k_0}\,  {\tilde{f}^{\text{i}}} _{02}\, e^{\sigma} & 
		-\frac{ i \, q_0 }{k_0} \, {\tilde{f}^{\text{i}}} _{02}  \,e^{-\sigma} &  {f^{\text{i}}} _{02}\, e^{-\rho}
	\end{vmatrix} = 0 \,,
\end{align}
where the expressions for the functions can be found in Eq.~\eqref{eqnodalx}.
The numerical roots, for the same parameter values as in Fig.~\ref{fig:nodalline_kx}, are shown in Fig.~\ref{fig:nodalline_boundstate}(b), represented by black dots.
Here, $ \sqrt{\left ({\mathcal M}-B \,k_{y}^2 \right )^2 + k_{z}^2}-V_0 \le E_b 
\le  \sqrt{ \left ( {\mathcal M}-B \,k_{y}^2 \right )^2 + k_{z}^2} $ is the inequality that $E_b$ obeys, which can be verified using Eq.~\eqref{energy_nodal}.

We find that the bound state landscape has changed significantly compared to the earlier case, since there are fewer and widely-spaced bound states for low values of $k_y$. The density of the bound states increases as $k_y$ crosses a threshold value. As before, the resonance points of Fig.~\ref{fig:nodalline_kx}(a) correspond to the red dots in Fig.~\ref{fig:nodalline_boundstate}(b), obeying $\efano-\hbar\omega = E_b$.

\subsection{Interpretation of the transport features from energies of bound states}

To corroborate the results in Figs.~\ref{fig:mweyl_kz} and \ref{fig:mweyl_kx}, we notice that $\mathcal A-V_0 \le E_b \le \mathcal A$ and $\mathcal B-V_0 \le E_b \le \mathcal B$ for wells oriented along the $z$-axis and $x$-axis, respectively, where $\mathcal A=\alpha_J \left(  k_x^2+k_y^2 \right)^{J/2}$ and
$\mathcal B = \left( \alpha_J^2 \, k_y^{2J} + v_z^2 \, k_z^2 \right)^{1/2}$. Now if we consider momentum values of the order of $8\,k_0$, noting that $(v_z/\alpha_J) = k_0^{J-1}$, we get
\begin{align}
\frac{\mathcal A} {\mathcal B} 
=  \frac{\lsb 1+ (k_x^2/k_y^2) \rsb^{J/2}}{\lsb 1+ k_0^{2(J-1)} \, 
(k_z^2/k_y^{2J}) \rsb^{1/2}} \sim \frac{2^{J/2}}{\lsb 1 + 8^{2 (1-J)} \rsb^{1/2}} \, .
\end{align}
For $J>1$, we find that $\mathcal A> \mathcal B$, which implies that bound states appear at relatively smaller energies for the $x$-oriented well. As a result, the corresponding Fano resonance points occur at lower energies, which is in agreement with the results.

\begin{figure}[] 
\centering
\subfigure[]{\includegraphics[width= 0.3\textwidth]{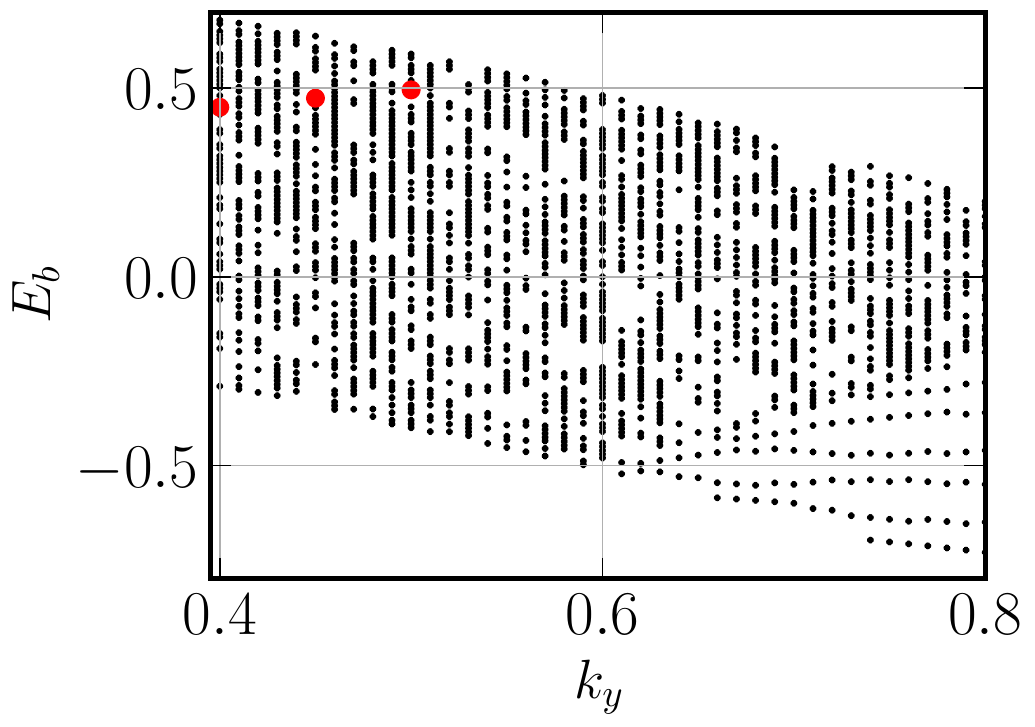}}\qquad
\subfigure[]{\includegraphics[width= 0.285 \textwidth]{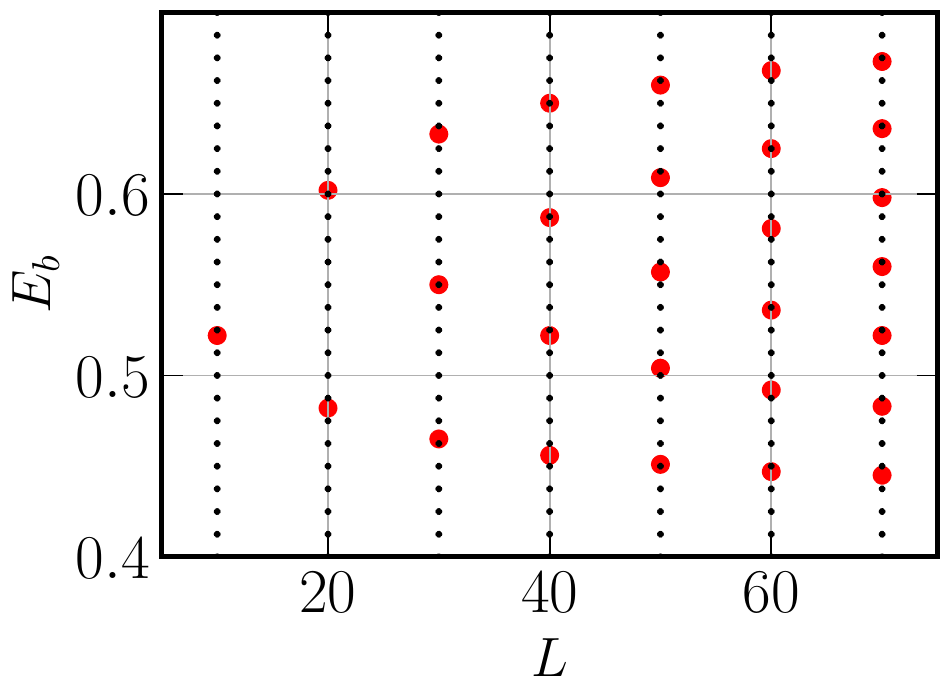}}\qquad
\subfigure[]{\includegraphics[width= 0.3 \textwidth]{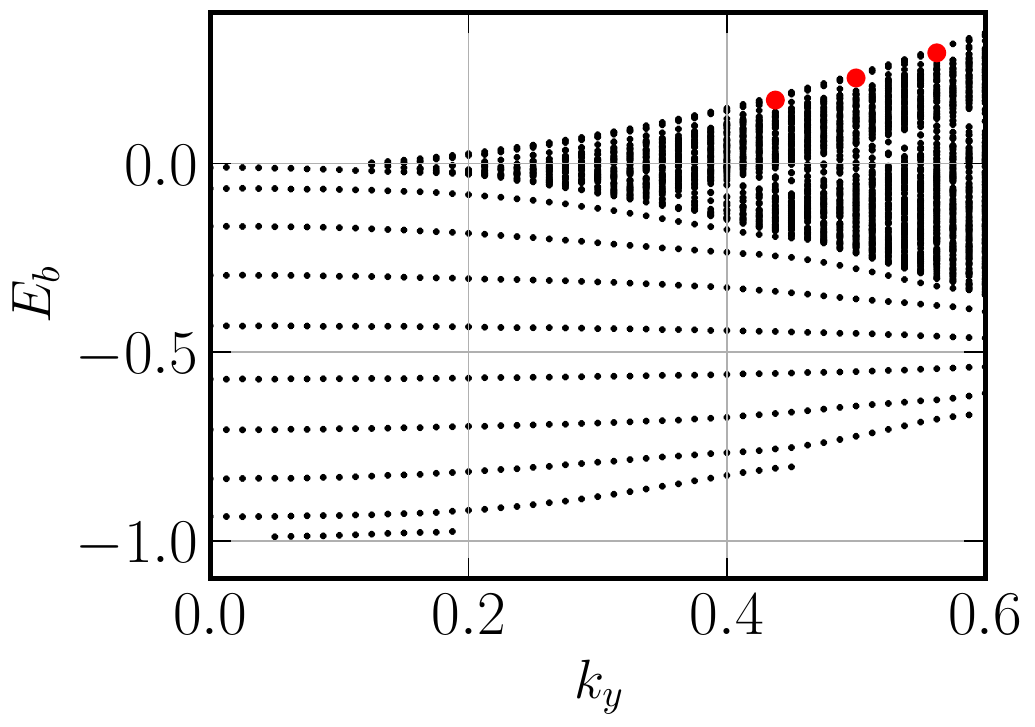}}
\caption{\label{fig:nodalline_boundstate}
Nodal-line semimetal: The bound states of a static quantum well, oriented along the $z$-direction [in (a) and (b)]; and $x$-direction [in (c)]. The red dots in all the subfigures represent the values of $E_b$ (in units of  ${\mathcal M}$). The red dots in (a) and (c) coincide with the energies of the Fano resonance points in Figs.~\ref{fig:nodalline_kz} and \ref{fig:nodalline_kx}, respectively, for the corresponding driven wells [with $V_1= \left( 1/40 \right) {\mathcal M}$ and $\hbar\omega= 0.25 \, {\mathcal M} $], such that $E_b = \efano -\hbar \omega$. The values of the parameters used are $V_0= {\mathcal M}$, $B =  {\mathcal M}^{-1}$, and $L= 15 \, {\mathcal M}^{-1} $. We have set $k_x = 0.4\,{\mathcal M} $ and $k_z=0$ for the plots in (a) and (c), respectively.
For (b), we have used $k_x=k_y = 0.4\,{\mathcal M}$.}
\end{figure}

Fig.~\ref{fig:nodalline_boundstate} shows that the number of bound states for the nodal-line semimetal is much higher if the well is oriented along the $z$-direction (compared to when it is aligned along the $x$-axis). This is the reason why there is no appreciable increase in the number of resonance points with an increase in $L$ for an $x$-oriented well. Fig.~\ref{fig:nodalline_boundstate}(b) illustrates why the successive Fano resonance points get denser as $L$ increases (thereby accommodating an extra Fano resonance point for an increase in $L$ by 10 units), as found in Fig.~\ref{fig:L_variation}.

\section{Summary and outlook} 
\label{sec:summary}

In this paper, we have investigated the effects of a periodic drive on the quantum mechanical transmission of quasiparticles, in 3D Weyl/multi-Weyl and nodal-line semimetals, through a potential well. We have used the Floquet scattering theory, focussing on the non-adiabatic limit. The periodic drive can be achieved by applying an ac gate voltage to the ends of a semiconducting structure, which creates a controlled and tunable time-dependent potential well. The oscillating well introduces equidistant Floquet side-bands in the dispersion. When one of the side-bands matches the energy of a (quasi)bound state of the quantum well, a sharp Fano resonance is observed in the transmission spectrum. The systems that we consider give rise to asymmetric resonance patterns (with extremely narrow bandwidths) with respect to the Fermi energy, which consists of a sharp dip followed by a sharp peak, or vice versa. This trend is similar to that in 2D electron gas/graphene \cite{Zhu15} and quadratic band-touching semimetals~\cite{Bera2021}, but in contrast with the symmetric resonances observed in pseudospin-1 Dirac-Weyl systems~\cite{Zhu17}. The presence of the Fano resonance points is also imprinted in the corresponding shot noise spectrum, where their signature is in the form of inflection points. The extremely narrow bandwidth resonances, that emerge in the semimetallic systems, can be potentially useful for designing advanced dynamic reconfigurable devices \cite{Karmakar19}. In experiments, these Fano resonances can be observed by measuring the pumped shot noise~\cite{shot_noise,shot_hg}, whose behaviour we have determined numerically.

For a Weyl/multi-Weyl semimetal, a well aligned along the $z$-direction leads to multiple Fano resonance points in the transmission spectrum. The number of resonances decreases for the anisotropic multi-Weyl cases, when the orientation of the well is changed to the $x$-axis. In both cases, the values of $\efano$ increase as $J$ increases.  

In case of the nodal-line semimetal, the resonance pattern shows a peak followed by a dip for a potential aligned along the $z$-axis, and a dip followed by a peak for a potential along the $x$-axis. Since the modulation in the shot noise is much weaker compared to the Weyl/multi‐Weyl cases, the behavior of the derivative of the shot noise with respect to $E_F$ helps us visualize the Fano resonance points better. 

The (quasi)bound states of the potential well are responsible for the emergence of the Fano resonances, and together with the Floquet side-band energy gap $\hbar \omega$, determine the values of $\efano$. Hence, $E_b$ can be computed via two complementary methods: (i) through the resonance condition $E_b=\efano-n \hbar\omega$, with $\efano$ obtained from the transmission curve; (ii) by solving for the energies of the bound states of the well explicitly. We have corroborated this relation for each case by considering the example of the bound states of the corresponding static potential well. In the parameter regime that we have considered, the matching occurs for the first ($n=1$) Floquet side-band, i.e., whenever the condition $E_b=\efano-\hbar\omega$ is satisfied.

For future research work, it will be interesting to look at the Floquet scattering properties in the presence of disorder \cite{rahul-sid,*ipsita-rahul,*ips-qbt-sc,ips-biref,ips-klaus} and/or magnetic fields \cite{mansoor,ips3by2,ips-aritra,ips-serena}. In the simplistic treatment pursued in this paper (and the related earlier papers \cite{ips-tunnel-qbcp,ips3by2,Zhu10,Zhu_2011,Dai2014,Bera2021}), we have considered single-particle Hamiltonians, and have ignored any electron‐electron or electron‐phonon interactions. Therefore, a complementary direction is to investigate the effects of the periodic potential in presence of interactions, which may introduce drastic effects like (i) destroying quantization of various physical quantities in the topological phases \cite{kozii,ips-photocurrent}; and/or (ii) emergence of strongly correlated phases~\cite{ips-seb,MoonXuKimBalents,rahul-sid,ipsita-rahul,ips-qbt-sc}, where quasiparticle description of transport turns out to be inapplicable \cite{ips-subir,ipsc2,ips-mem-mat,*ips-freire2,*ips-freire3}.

\section*{Acknowledgments}
S.S. is funded by the National Science Centre (Narodowe Centrum Nauki), Poland, under the scheme Preludium Bis-2 (Grant Number 2020/39/O/ST3/00973).

\appendix

\section{{S-matrix for various cases}} 
\label{appA}

In Eq.~\eqref{eq:psi}, there are six unknown coefficients: $A^{\t{i}}_n$ and $A^{\t{o}}_n$ for region-1, $\alpha_m$ and $\beta_m$ for region-2, and $B^{\t{i}}_n$ and $B^{\t{o}}_n$ for region-3. 
By using the continuity of the wavefunction at the boundaries $a=\pm L/2$, we get the expressions
\begin{align}
	&A_{n}^{\t{i}}(t) \,e^{-i\,k_{n}L/2}
	\begin{pmatrix}
		 {f^{\text{i}}} _{n1}  \\  {f^{\text{i}}} _{n2}  \\
	\end{pmatrix}
	+\,A_{n}^{\t{o}}(t) \,e^{i \,k_{n}L/2}
	\begin{pmatrix}
		 {f^{\text{o}}} _{n1}  \\  {f^{\text{o}}} _{n2}  \\
	\end{pmatrix} 
	= 
	\sum \limits _{m=-\infty}^\infty \left[   
	\alpha_{m} (t)\,e^{-i \,q_{m} L/2}
	\begin{pmatrix}
		 {\tilde{f}^{\text{i}}} _{m1}  \\  {\tilde{f}^{\text{i}}} _{m2}  \\
	\end{pmatrix}
	+  \beta_{m} (t)\,e^{i \,q_{m} L/2}
	\begin{pmatrix}
		 {\tilde{f}^{\text{o}}} _{m1}  \\
		 {\tilde{f}^{\text{o}}} _{m2}  \\
	\end{pmatrix} \right ],\nn
	\nn
	& \sum \limits _{m=-\infty}^\infty \left[   \alpha_{m} (t)\,e^{i \,q_{m}L/2}
	\begin{pmatrix}
		 {\tilde{f}^{\text{i}}} _{m1}  \\  {\tilde{f}^{\text{i}}} _{m2}  \\
	\end{pmatrix}
	+  \beta_{m} (t)\,e^{-i \,q_{m}L/2}
	\begin{pmatrix}
		 {\tilde{f}^{\text{o}}} _{m1}  \\  {\tilde{f}^{\text{o}}} _{m2}  \\
	\end{pmatrix} \right ] =	B_{n}^{i} (t)\,e^{-i \,k_{n}L/2}
	\begin{pmatrix}
		 {f^{\text{o}}} _{n1}  \\
		 {f^{\text{o}}} _{n2}  \\
	\end{pmatrix} 
	+ B_{n}^{o}(t) \,e^{i \,k_{n}L/2}
	\begin{pmatrix}
		 {f^{\text{i}}} _{n1}  \\
		 {f^{\text{i}}} _{n2}  \\
	\end{pmatrix}, 
\end{align}
where we have restricted to the case $E_m>-V_0$ in order to avoid cluttering.
We parametrize these relations as:
\begin{align}
	& A^{\t{o}}_n   = \sum_m 
	\lrb v_{11}^{nm} \, \alpha_m + v_{12}^{nm} \, \beta_m\rrb, \quad
	B^{\t{o}}_n  = \sum_m 
	\lrb v_{21}^{nm} \, \alpha_m + v_{22}^{nm} \, \beta_m\rrb, \nn
	A^{\t{i}}_n \, e^{-i \, k_n\, L/2} & = 
	\sum_m 
	\lrb u_{11}^{nm} \, \alpha_m + u_{12}^{nm} \, \beta_m\rrb, \quad
	B^{\t{i}}_n \, e^{-i \, k_n \,L/2}  = \sum_m 
	\lrb u_{21}^{nm} \, \alpha_m + u_{22}^{nm} \, \beta_m\rrb,
\end{align}
which are now rewritten in a compact form in terms of matrices as follows:
\begin{align}
	\label{eqbc}
	& {\mathbf A}^{\t{o}}=   \boldsymbol {v}_{11} \cdot \bm{\alpha}
	+\boldsymbol {v}_{12} \cdot \bm{\beta} \,, \quad
	\mathbf {B}^{\t{o}}=   \boldsymbol {v}_{21} \cdot \bm{\alpha}+\boldsymbol {v}_{22} \cdot \bm{\beta} \,,\quad
	\mathbf {M}_{r} \cdot \mathbf {A}^{\t{i}}
	= {\boldsymbol{u}}_{11} \cdot \bm{\alpha}+{\boldsymbol{u}}_{12} \cdot \bm{\beta}\,,\quad
	\mathbf {M}_{r} \cdot \mathbf {B}^{\t{i}}r
	= {\boldsymbol{u}}_{21} \cdot \bm{\alpha} + {\boldsymbol{u}}_{22} \cdot \bm{\beta}\,,
\end{align}
\begin{align}
	[\mathbf {M}_{r} ]_{nm}= e^{- i \,k_{n} L / 2} \,\delta_{n,m}\,.
\end{align}
The coefficients of $\alpha_m$ and $\beta_m$ depend on the Hamiltonian under consideration, and also on the orientation of the potential well. We will provide their explicit expressions on a case-by-case basis in the following subsections.

Let us now define the matrices
\begin{align}
	\mathbf {M}_{AA}
	& =\frac{ \boldsymbol {v}_{11} \cdot {\boldsymbol{u}}^{-1}_{12}\cdot \mathbf {M}_{r}}
	{{\boldsymbol{u}}^{-1}_{12} \cdot {\boldsymbol{u}}_{11}
		-{\boldsymbol{u}}^{-1}_{22} \cdot {\boldsymbol{u}}_{21}}
	+\frac{ \boldsymbol {v}_{12} \cdot {\boldsymbol{u}}^{-1}_{11}\cdot \mathbf {M}_{r}}
	{{\boldsymbol{u}}^{-1}_{11} \cdot {\boldsymbol{u}}_{12}-{\boldsymbol{u}}^{-1}_{21} \cdot {\boldsymbol{u}}_{22}}\,, \quad
	\mathbf {M}_{AB}=-\frac{ \boldsymbol {v}_{11} \cdot {\boldsymbol{u}}^{-1}_{22}\cdot \mathbf {M}_{r}}{{\boldsymbol{u}}^{-1}_{12} \cdot {\boldsymbol{u}}_{11}-{\boldsymbol{u}}^{-1}_{22} \cdot {\boldsymbol{u}}_{21}}-\frac{ \boldsymbol {v}_{12} \cdot {\boldsymbol{u}}^{-1}_{21}\cdot \mathbf {M}_{r}}{{\boldsymbol{u}}^{-1}_{11} \cdot {\boldsymbol{u}}_{12}-{\boldsymbol{u}}^{-1}_{21} \cdot {\boldsymbol{u}}_{22}} \,, \nn
	\mathbf {M}_{BA}
	& =\frac{ \boldsymbol {v}_{21} \cdot {\boldsymbol{u}}^{-1}_{12}\cdot \mathbf {M}_{r}}{{\boldsymbol{u}}^{-1}_{12} \cdot {\boldsymbol{u}}_{11}-{\boldsymbol{u}}^{-1}_{22} \cdot {\boldsymbol{u}}_{21}}+\frac{ \boldsymbol {v}_{22} \cdot {\boldsymbol{u}}^{-1}_{11}\cdot \mathbf {M}_{r}}{{\boldsymbol{u}}^{-1}_{11} \cdot {\boldsymbol{u}}_{12}
		-{\boldsymbol{u}}^{-1}_{21} \cdot {\boldsymbol{u}}_{22}} \,, \quad
	\mathbf {M}_{BB}=-\frac{ \boldsymbol {v}_{21} \cdot {\boldsymbol{u}}^{-1}_{22}\cdot \mathbf {M}_{r}}{{\boldsymbol{u}}^{-1}_{12} \cdot {\boldsymbol{u}}_{11}-{\boldsymbol{u}}^{-1}_{22} \cdot {\boldsymbol{u}}_{21}}-\frac{ \boldsymbol {v}_{22} \cdot {\boldsymbol{u}}^{-1}_{21}\cdot \mathbf {M}_{r}}{{\boldsymbol{u}}^{-1}_{11} \cdot {\boldsymbol{u}}_{12}-{\boldsymbol{u}}^{-1}_{21} \cdot {\boldsymbol{u}}_{22}} \,,
\end{align} 
which help us to write the relations in Eq.~\eqref{eqbc} as
\begin{align}
	\label{eqSmatrix0}
	\pmtx{\mathbf{A}^{\t o} \\ \mathbf{B}^{\t o}} 
	= \boldsymbol{\mathcal S}\, \pmtx{\mathbf{A}^{\t i} \\ \mathbf{B}^{\t i}},\quad
	\boldsymbol{\mathcal S} = 
	\pmtx{\mathbf{M}_{AA} & \mathbf{M}_{AB} \\ \mathbf{M}_{BA} & \mathbf{M}_{BB}} .
\end{align}
Hence, for the $n^{\rm{th}}$ Floquet side-band, we get
\begin{align}
	\label{eqSmatrix1}
	\pmtx{ {A}^{\t o}_n \\  {B}^{\t o}_n } 
	= \sum \limits_{m=-\infty}^{\infty}
	\mathcal{S}_{nm}\, \pmtx{ {A}^{\t i}_m \\  {B}^{\t i}_m }.
\end{align}
Note that $\mathcal{S}_{nm}$ represents the $(n,m)$ component of $\boldsymbol{\mathcal S} $, and thus $\mathcal{S}_{nm}$ itself is a matrix composed of the $(n,m)$ components of $\mathbf{M}_{AA}$, $\mathbf{M}_{AB}$, $\mathbf{M}_{BA}$, and $\mathbf{M}_{BB}$.
The matrix $s$ in Eq.~\eqref{eqsmatrix} can now be determined from $\mathcal{S}_{nm}$.

Let us now show the explicit expressions for $\boldsymbol {v}_{11}$, $\boldsymbol {v}_{12}$, $\boldsymbol {v}_{21}$,
$\boldsymbol {v}_{22} $, $\boldsymbol {u}_{11}$, $\boldsymbol {u}_{12}$, $\boldsymbol {u}_{21}$, and
$\boldsymbol {u}_{22} $ for the first three set-ups studied in the main text.

\subsection{Weyl/multi-Weyl semimetal with potential well aligned along the $z$-axis}
\label{appweylz}

For a quasiparticle in a Weyl/multi-Weyl semimetal, propagating through a potential well aligned along the $z$-axis, the components of the matrices take the forms:
\begin{align}
	v_{11}^{nm} & = \frac{n_2 \,e^{-i \, (k_n+q_m)L/2}} {2 \,n_3}  
	\lrb 1 + \frac{\zeta_n}{k_n}-\frac{\zeta_m}{k_n} -\frac{q_m}{k_n}\rrb   \jbessel, 
	\,\,
	v_{12}^{nm} = \frac{n_2 \,e^{-i \, (k_n-q_m)L/2}} {2 \,n_4}   
	\lrb 1 + \frac{\zeta_n}{k_n}-\frac{\zeta_m}{k_n} +\frac{q_m}{k_n}\rrb  \jbessel, \nn
	v_{21}^{nm} &= \frac{n_1 \,e^{-i \, (k_n-q_m)L/2}} {2 \,n_3}    
	\lrb  1-\frac{\zeta_n}{k_n}+\frac{\zeta_m}{k_n}+\frac{q_m}{k_n}\rrb  \jbessel, \,\,
	v_{22}^{nm}  = \frac{n_2 \,e^{-i \, (k_n+q_m)L/2}} {2 \,n_4}  
	\lrb 1-\frac{\zeta_n}{k_n}
	+\frac{\zeta_m}{k_n} -\frac{q_m}{k_n}\rrb  \jbessel, 
\end{align}
and
\begin{align}	
	u_{11}^{nm} &= \frac{n_1 \,e^{-i \,  q_m L/2}} {2 \,n_3}    \lrb 
	1 -\frac{\zeta_n}{k_n}+\frac{\zeta_m}{k_n} +\frac{q_m}{k_n}\rrb  \jbessel, \,\,
	u_{12}^{nm} = \frac{n_1 \,e^{i \,  q_m L/2}} {2 \,n_4}    
	\lrb 
	1-\frac{\zeta_n}{k_n}+\frac{\zeta_m}{k_n}-\frac{q_m}{k_n}\rrb  \jbessel, \nn
	u_{21}^{nm} &= \frac{n_2 \,e^{i \,  q_m L/2} } {2 \,n_3}   
	\lrb 1 + \frac{\zeta_n}{k_n}
	-\frac{\zeta_m}{k_n} -\frac{q_m}{k_n}\rrb  \jbessel, \,\,
	u_{22}^{nm} = \frac{n_2 \,e^{-i \,  q_m L/2}} {2 \,n_4}    
	\lrb 1 + \frac{\zeta_n}{k_n}-\frac{\zeta_m}{k_n} +\frac{q_m}{k_n}\rrb  \jbessel \, .
\end{align}

\subsection{Multi-Weyl semimetal with potential well aligned along the $x$-axis}
\label{appweylx}

For a quasiparticle in a Weyl/multi-Weyl semimetal, propagating through a potential well aligned along the $x$-axis, the components of the matrices take the forms:
\begin{align}
	v_{11}^{nm} &= \frac{n_1 \, e^{-i \, (k_n+q_m)L/2}
		\lsb (\rynp)^J - (\rymp)^J E_{nm} \rsb
	}
	{n_3 \lsb (\rynp)^J- (\rynm)^J \rsb}  
	\,  \jbessel, \,\,
	v_{12}^{nm} = \frac{n_1 \, e^{-i \, (k_n-q_m)L/2}}{n_3 \lsb (\rynp)^J- (\rynm)^J \rsb}   \lsb (\rynp)^J - (\rymm)^J E_{nm} \rsb \jbessel, \nn
	v_{21}^{nm} &= \frac{n_1 \, e^{-i \, (k_n-q_m)L/2}
		\lsb  (\rymp)^J E_{nm} -(\rynm)^J  \rsb 
	}
	{n_3 \lsb (\rynp)^J- (\rynm)^J \rsb} \, \jbessel, \,\,
	v_{22}^{nm} = \frac{n_1 \, e^{-i \, (k_n+q_m)L/2}
		\lsb  (\rymm)^J E_{nm} -(\rynm)^J \rsb
	}
	{n_3 \lsb (\rynp)^J- (\rynm)^J \rsb}   
	\,\jbessel,
\end{align}
and
\begin{align}	
	u_{11}^{nm} &= \frac{n_1 \, e^{-i \,  q_m L/2} \lsb -(\rynm)^J + (\rymp)^J E_{nm} \rsb }
	{n_3 \lsb (\rynp)^J- (\rynm)^J \rsb}  
	\,\jbessel, \,\,
	u_{12}^{nm} = \frac{n_1 \, e^{i \,  q_m L/2}
		\lsb  (\rymm)^J E_{nm} -(\rynm)^J \rsb
	}
	{n_3 \lsb (\rynp)^J- (\rynm)^J \rsb}   \, \jbessel, \nn
	u_{21}^{nm} &= \frac{n_1 \, e^{i \,  q_m L/2} \lsb (\rynp)^J - (\rymp)^J E_{nm} \rsb
	}
	{n_3 \lsb (\rynp)^J- (\rynm)^J \rsb}   
	\, \jbessel, \,\,
	u_{22}^{nm} = \frac{n_1 \, e^{-i \,  q_m L/2}  \lsb (\rynp)^J - (\rymm)^J E_{nm} \rsb
	}
	{n_3 \lsb (\rynp)^J- (\rynm)^J \rsb}  \, \jbessel,
\end{align}
where
\begin{align}
	r_{yn}^{\pm} = k_y \pm i \, k_n\,, \quad 
	r_{ym}^{\pm} = k_y\pm i \,  q_m \,, \quad 
	E_{nm} = \frac{E_n-k_z}{E_m+V_0-k_z} \, .
\end{align}

\subsection{Nodal-line semimetal with potential well aligned along the $z$-axis}
\label{appnodalz}

For a quasiparticle in a nodal-line semimetal, propagating through a potential well aligned along the $z$-axis, the components of the matrices take the forms:
\begin{align}
	v_{11}^{nm} & = \frac{n_2 \,e^{-i \, (k_n+q_m)L/2}} {2 \,n_3}  
	\lrb \frac{k_n-V_0-q_m}{k_n} \rrb  \jbessel, \,\,
	v_{12}^{nm} = \frac{n_2 \,e^{-i \, (k_n-q_m)L/2} } {2 \,n_4}  
	\lrb \frac{k_n-V_0+q_m}{k_n} \rrb  \jbessel,
	\nn
	v_{21}^{nm} &= \frac{n_1 \,e^{-i \, (k_n-q_m)L/2}} {2 \,n_3}    
	\lrb \frac{k_n+V_0+q_m}{k_n} \rrb  \jbessel, \,\,
	v_{22}^{nm} = \frac{n_1 \,e^{-i \, (k_n+q_m)L/2}} {2 \,n_4}    
	\lrb \frac{k_n+V_0-q_m}{k_n} \rrb  \jbessel,
\end{align}
and
\begin{align}
	u_{11}^{nm} & = \frac{n_1 \,e^{-i \,  q_m L/2} } {2 \,n_3}   
	\lrb \frac{k_n+V_0+q_m}{k_n} \rrb  \jbessel, \,\,
	u_{12}^{nm} = \frac{n_1  \,e^{i \,  q_m L/2}} {2 \,n_4} 
	\lrb \frac{k_n+V_0-q_m}{k_n} \rrb  \jbessel, \nn
	u_{21}^{nm} & = \frac{n_2  \,e^{i \,  q_m L/2}} {2 \,n_3}   
	\lrb \frac{k_n-V_0-q_m}{k_n} \rrb  \jbessel, \,\,
	u_{22}^{nm} = \frac{n_2 \,e^{-i \,  q_m L/2}} {2 \,n_4} 
	\lrb \frac{k_n-V_0+q_m}{k_n} \rrb  \jbessel  .
\end{align}

\section{S-matrix for nodal-line semimetal with potential well aligned along the $x$-axis}
\label{appnlx}

When a quasiparticle in a nodal-line semimetal is propagating through a potential well aligned along the $x$-axis, we have the functions defined in Eq.~\eqref{eqnodalx}. Again, we restrict to the case $E_m>-V_0$ in order to avoid cluttering. For this scattering problem, we find that $ {f^{\text{o}}} _{n1} =  {f^{\text{i}}} _{n1}$,
$  {f^{\text{o}}} _{n2} =  {f^{\text{i}}} _{n2}$, $ {\tilde{f}^{\text{o}}} _{m1} =  {\tilde{f}^{\text{i}}} _{m1} $, and $ {\tilde{f}^{\text{o}}} _{m2} =  {\tilde{f}^{\text{i}}} _{m2}$, which leads to some simplification. But, in this case, we need to use the boundary conditions resulting from matching both the wavefunction and its derivative along the $x$-axis.
To express these relations, it is convenient to introduce the matrices $\mathbf{M}^\pm$, $\mathbf{M}_i$, $\mathbf {C}^\pm$,
$\mathbf{M}_r$, and $\mathbf{M}_c^\pm$, whose components are given by
\begin{align}
	& \left[{ M}^{\pm} \right ]_{nm}
	=   \frac{ {\tilde{f}^{\text{i}}} _{m2}}  {  {f^{\text{i}}} _{n2}} 
	\left[ \left (1+\frac{q_{m}} {k_n} \right ) 
	\,e^{- i \,q_{m} L/ 2 } \pm    
	\left (1-\frac{q_{m}}{k_n} \right ) \,
	e^{  i \,q_{m} L/2} \right]   J_{n-m} \bigg( \frac{V_{1}}{\hbar \omega}\bigg) \,,
	\quad \left[ M_{i}\right]_{nm}= e^{-i \, k_n L}\, \delta_{n,m}\,,
	\nn & 
	\quad \left[ C \right ] _{m}^{\pm}= \alpha_{m} \pm \beta_{m}\,,\quad
	\left[ M_{r} \right] _{nm}= 2\,e^{-i \,k_n L/2 } \,\delta_{n,m}\,,\quad 
	\left[{   M_{c}}^{\pm} \right]_{nm}= 
	\frac{  {\tilde{f}^{\text{i}}} _{m2}} {  {f^{\text{i}}} _{n2}} \,
	e^{-  i\, \left (k_n \pm q_{m} \right ) L/2}\,
	J_{n-m}\bigg ( \frac{V_{1}}{\hbar \omega}\bigg) \,.
\end{align}
This helps us to write the solutions for the coefficients in a compact form as follows:
\begin{align}
	\mathbf{M}_r  \cdot
	\left 	(\mathbf{A}^{\t i} \pm \mathbf{B}^{\t i} \right )
	& = \mathbf{M} ^\pm \cdot \mathbf{C}^\pm 
	\Rightarrow 
	\*C^{\pm} = (\*M ^{\pm})^{-1} \cdot \*M_r \cdot(\*A^{\t i} \pm \*B^{\t i})\,,\nn
	\mathbf{A}^{\t o} &= \mathbf{M}_c^{+} \cdot
	\lrb \frac{\mathbf{C}^+ + \mathbf{C}^-}{2} \rrb 
	+ \mathbf{M}_c^{-} \cdot
	\lrb \frac{\mathbf{C}^+ - \mathbf{C}^-}{2}  \rrb 
	- \mathbf{M}_i  \cdot
	\mathbf{A}^{\t i} \equiv 
	\mathbf{M}_{AA}  \cdot \mathbf{A}^{\t i} 
	+ \mathbf{M}_{AB} \cdot \mathbf{B}^{\t i}\,, \nn
	\mathbf{B}^{\t o} &= \mathbf{M}_c^{-} \cdot
	\lrb \frac{\mathbf{C}^+ - \mathbf{C}^-}{2} \rrb 
	+ \mathbf{M}_c^{+} \cdot
	\lrb \frac{\mathbf{C}^+ 
		+ \mathbf{C}^-}{2}  \rrb 
	- \mathbf{M}_i \cdot \mathbf{B}^{\t i} \equiv \mathbf{M}_{BA} \cdot\mathbf{A}^{\t i} 
	+ \mathbf{M}_{BB} \cdot\mathbf{B}^{\t i}\,,
\end{align}
using the fact that
\begin{align}
	\*C^{+} +\*C^{-}  &= 
	\lsb (\*M ^+)^{-1}+(\*M^-)^{-1}  \rsb \cdot \*M_r\cdot \mathbf{A}^{\t i} 
	+ \lsb  (\*M ^+)^{-1}-(\*M ^-)^{-1}  \rsb \cdot \*M_r \cdot \mathbf{B}^{\t i} \,,\nn
	\*C^{+} -\*C^{-}  
	& = \lsb  (\*M ^+)^{-1}-(\*M ^-)^{-1}  \rsb \cdot  \*M_r \cdot \mathbf{A}^{\t i} 
	+ \lsb  (\*M ^+)^{-1}+(\*M ^-)^{-1}  \rsb \cdot \*M_r \cdot\mathbf{B}^{\t i}\,.
\end{align}
The final expression can now be formulated in terms of the matrices shown in Eqs.~\eqref{eqSmatrix0} and \eqref{eqSmatrix1}, leading to
the matrix $s$ in Eq.~\eqref{eqsmatrix}.


\bibliography{ref}

\end{document}